\documentclass[a4paper,
               biblatex,     
               ]{jacow}
\usepackage{pdfpages,multirow,ragged2e} %
\usepackage{transparent}
\makeatletter%
	\ifboolexpr{bool{xetex}}
	 {\renewcommand{\Gin@extensions}{.pdf,%
	                    .png,.jpg,.bmp,.pict,.tif,.psd,.mac,.sga,.tga,.gif,%
	                    .eps,.ps,%
	                    }}{}
\makeatother

\ifboolexpr{bool{xetex} or bool{luatex}} 
 {}                                      
 {\usepackage[utf8]{inputenc}}           

\usepackage{subfigure}
\usepackage{graphicx}
\usepackage{siunitx}
\usepackage{wasysym}

\usepackage{amsmath}
\begin{document}

\title{Correlation of srf performance to oxygen diffusion length of medium temperature heat treated cavities\thanks{This work was funded by the Helmholtz Association within the MT ARD and the European XFEL R\&D Program.}}

\author{C. Bate\thanks{christopher.bate@desy.de}, K. Kasprzak, D. Reschke,\\
L. Steder, L. Trelle, H. Weise, M. Wiencek, J. Wolff \\
Deutsches Elektronen-Synchrotron DESY, Germany \\
  }
	
\maketitle

\begin{abstract}

This comprehensive study, being part of the European XFEL R\&D effort, elucidates the influence of medium temperature (mid-T) heat treatments between 250°C and 350°C on the performance of 1.3~GHz superconducting radiofrequency (SRF) niobium cavities. 
Utilizing a refurbished niobium retort furnace equipped with an inter-vacuum chamber and cryopumps at DESY, we have embarked on an investigation to enhance the state-of-the-art SRF cavity technology. Our research reveals that mid-T heat treatments significantly boost the quality factor ($Q_0$) of the cavities, achieving values between $2\cdot10^{10}$ to $5\cdot10^{10}$ at field strengths around 16~MV/m, while the maximum field strengths are limited to 25-35~MV/m and enhanced sensitivity to trapped magnetic flux is observed. Moreover, we delve into the effects of surface impurity concentration changes, particularly the diffusion of oxygen content, and its impact on performance enhancements. By categorizing treatments based on calculated diffusion lengths using the whole temperature profile, we recognize patterns that suggest an optimal diffusion length conducive to optimizing cavity performance. SIMS results from samples confirm the calculated oxygen diffusion lengths in most instances. Deviations are primarily attributed to grain boundaries in fine-grain materials, necessitating repeated measurements on single-crystal materials to further investigate this phenomenon. Investigations into cooling rates and the resulting spatial temperature gradients across the cavities ranging from 0.04 to 0.2~K/mm reveal no significant correlation with performance following a mid-T heat treatment. However, the increased sensitivity to trapped magnetic flux leads to new challenges in the quest for next-generation accelerator technologies since the requirement for magnetic hygiene gets stricter. 
\end{abstract}

\section{The mid-T heat treatment recipe for niobium srf cavities}

In-situ medium-temperature bake experiments at 250-400°C \cite{SamMidT} revealed outstanding quality factors on 1.3~GHz single-cell cavities. Subsequent studies \cite{He_2021,HIto} employed commercially available ultra-high vacuum (UHV) furnaces for mid-T heat treatments, followed by sequential cavity cleaning and assembly in a clean-air environment.\\
The recipes we aim to investigate here involve baking the cavity in UHV at 250-350°C for a constant duration between three to twenty hours \cite{steder:linac2022-thpoge22, bate:srf2023-mopmb022}.
%
%
%
A key advantage of mid-T heat treatment is the elimination of the need for post-heating chemical surface treatments. Additionally, the recipe is highly reproducible, it offers significantly shorter baking times and eliminates the requirement for additional gases, such as nitrogen, in the furnace.

\section{The Niobium Retort Furnace at DESY}

\begin{figure}[!htb]
   \centering
   \includegraphics*[width=0.8\columnwidth]{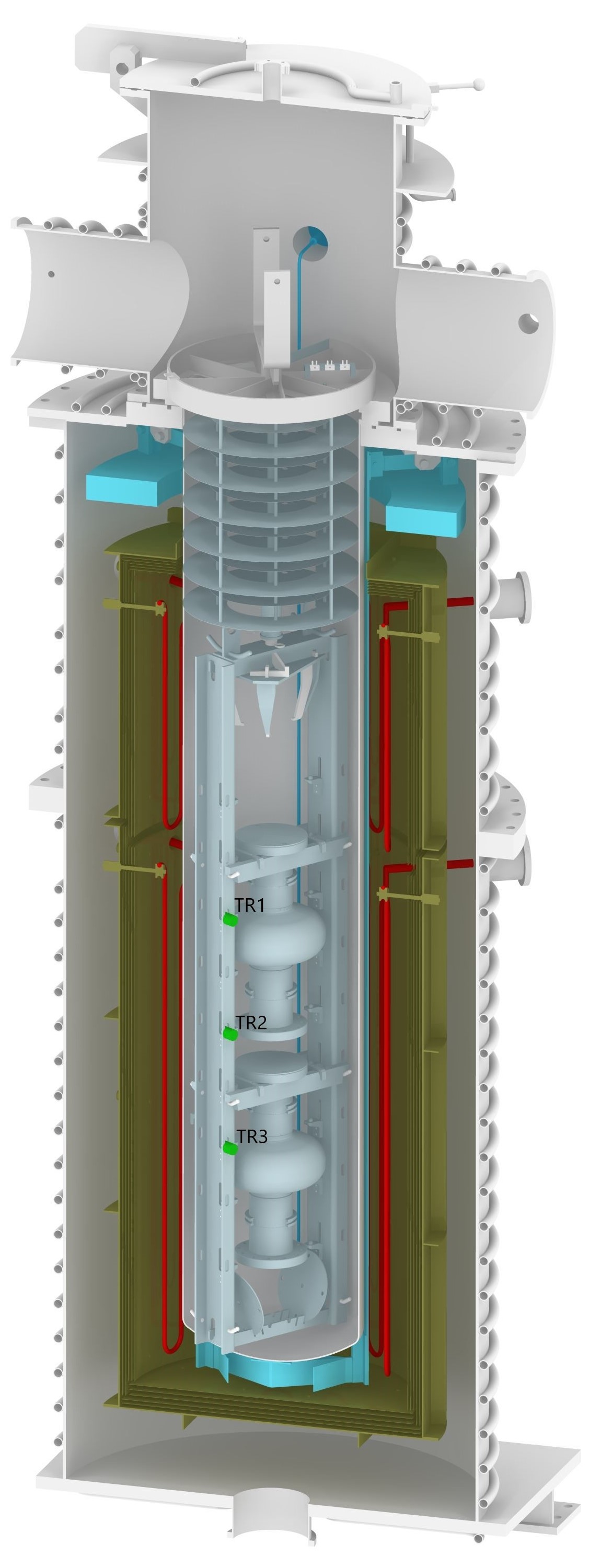}
   \caption{Cross-sectional depiction of the niobium furnace, presenting the furnace installation frame in which the cavities are installed and thermal shielding of the retort in grey-blue, heaters in red, thermal shielding within the support vacuum in yellow, and the surrounding vessel with water-cooling in light grey. The design includes a retort support structure to counteract deformations at high temperatures due to the softening of the material which is shown in cyan. The placement of three temperature sensors labelled TR1, TR2, and TR3, on the furnace frame, which serves as the mounting platform for the installed cavities, is shown.}
   \label{fig:furnacemodel}
\end{figure}
\begin{figure}[!htb]
   \centering
   \includegraphics*[width=1.0\columnwidth]{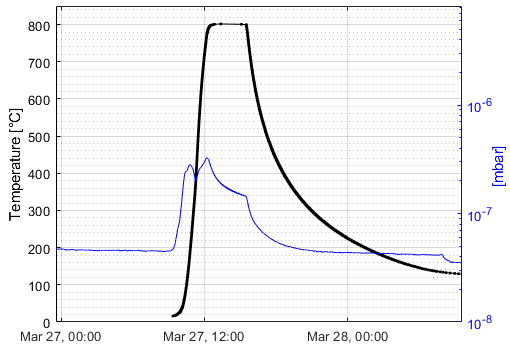}
   \caption{Time evolution plot depicting temperature and pressure during a 3-hour run at 800°C.}
   \label{fig:800Ctestrun}
\end{figure}

The furnace, designed for vertical loading of 1.3~GHz nine-cell cavities, was originally built in 1992 for titanium post-purification \cite{padamsee2008rf}. 
Situated in the ISO 4 area of the cavity assembly clean room, the furnace features a niobium retort with a separate support vacuum housing the heaters. With a usable diameter of $0.3~\text{m}$ and a depth of $1.3~\text{m}$, it can be used to bake a $1.3~\text{GHz}$ nine-cell cavity or up to two single-cell cavities positioned vertically above each other (see Fig.~\ref{fig:furnacemodel}) which we refer to as so-called "tandem run". In all cases the cavities are equipped with a niobium cap on the upper flange for particle protection. More details can be found in \cite{LennartSRF23}.\\
%
%
A complete furnace refurbishment was conducted, which included renewing the vacuum, cooling, and control systems and implementing partial pressure control and a mass spectrometry system. The oil-free cryo pumping system supports a base pressure of $2\cdot10^{-8}$~mbar at room temperature. During cavity treatment at 800°C, the pressure gradually rises to approximately 2-$3\cdot10^{-7}$~mbar as shown in Fig.~\ref{fig:800Ctestrun}.\\
%
%
Ongoing efforts focus on refining furnace control features and establishing reproducible treatment protocols. Plans include treating a nine-cell cavity with the mid-T heat treatment, aiming to develop industrial application recipes. 
%
\subsection{Temperature monitoring and controlling}
\begin{figure}[!htb]
   \centering
   \includegraphics*[width=1.0\columnwidth]{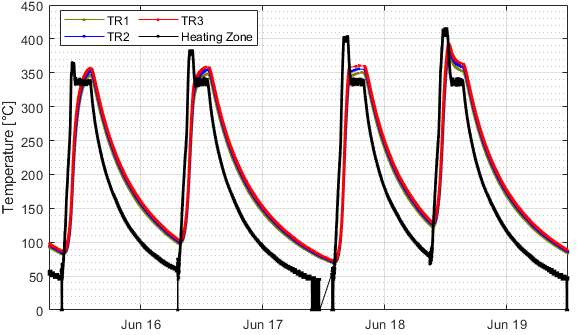}
   \caption{Time evolution plot depicting the temperatures measured by three temperature sensors (TR1-3), along with the temperature of a designated heater employed for controlling and control optimization of the heating zone. The plot illustrates the control optimization process during a 350°C 3-hour treatment, showcasing the intentional overshooting of the heater temperature to achieve the desired temperature profile (third ramping curve) at the cavities.}
   \label{fig:overshoot}
\end{figure}

%
%
The furnace is designed to achieve temperatures of up to 1400°C. Two tungsten heating coils, positioned in the upper and lower sections, are situated within the support vacuum alongside the thermal shielding.\\
Three type-S thermocouples are used to monitor the temperature inside the recipient. These sensors are attached to the furnace installation frame made of niobium, holding the cavities. The measuring tips are each inserted in dedicated small niobium blocks at the frame. The sensor positions are shown in Fig.~\ref{fig:furnacemodel}. This installation was considered to be the most reproducible and secure method to determine reasonable and consistent temperature values. 
%
%
Since the furnace comprises a two-stage vacuum as shown in Fig.~\ref{fig:furnacemodel}, and the heating elements are located in the support vacuum, the thermal response is delayed and control optimization was essential. 
Due to this design it is especially challenging to control temperatures below 350°C. 
Currently, the solution is to implement a slight "overshoot" to achieve the desired flat temperature profile as shown in Fig.~\ref{fig:overshoot} where it was adapted for the 3-hours 350°C treatment. As seen, in the first two curves of Fig.~\ref{fig:overshoot}, the profile is still too sharp, and it is only from the third ramp-up that the profile begins to approximate a constant at 350°C for 3 hours. In the fourth ramp-up, it can already be observed that the overshoot was too high, causing the profile to exceed 350°C. Overshooting is not necessary for higher temperatures such as 800°C due to the higher thermal conductivity of the system.



\section{Experimental overview}

\begin{figure}[!htb]
   \centering
   \includegraphics*[width=1.0\columnwidth]{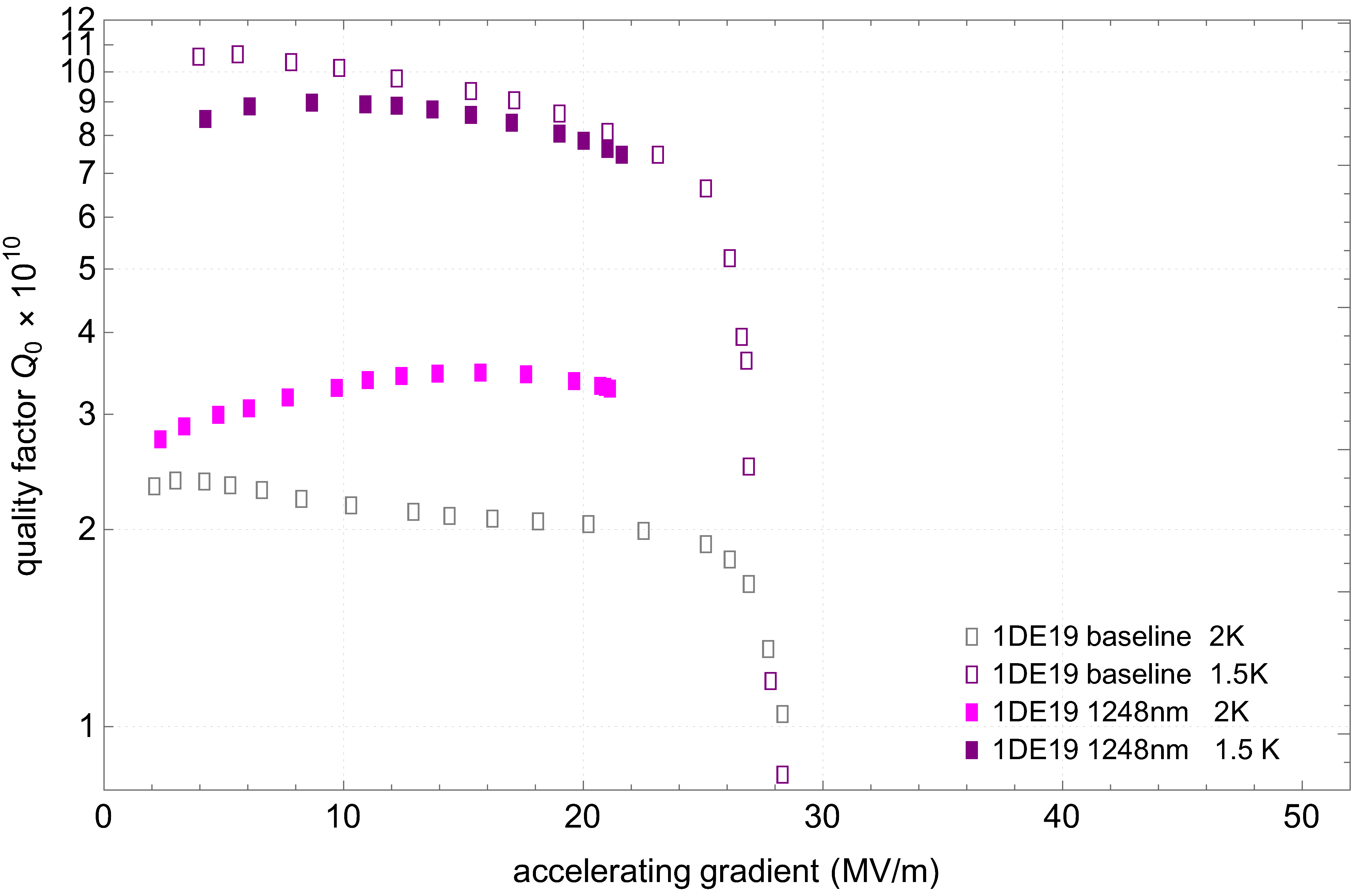}
   \caption{\(Q(E)\) at 1.5~K and 2~K before (baseline) and after a mid-T heat treatment of 4.5 h at 335°C corresponding to a calculated oxygen diffusion length of 1248~nm.}
   \label{fig:1DE19withbaseline}
\end{figure}
\begin{figure}[!htb]
   \centering
   \includegraphics*[width=0.9\columnwidth]{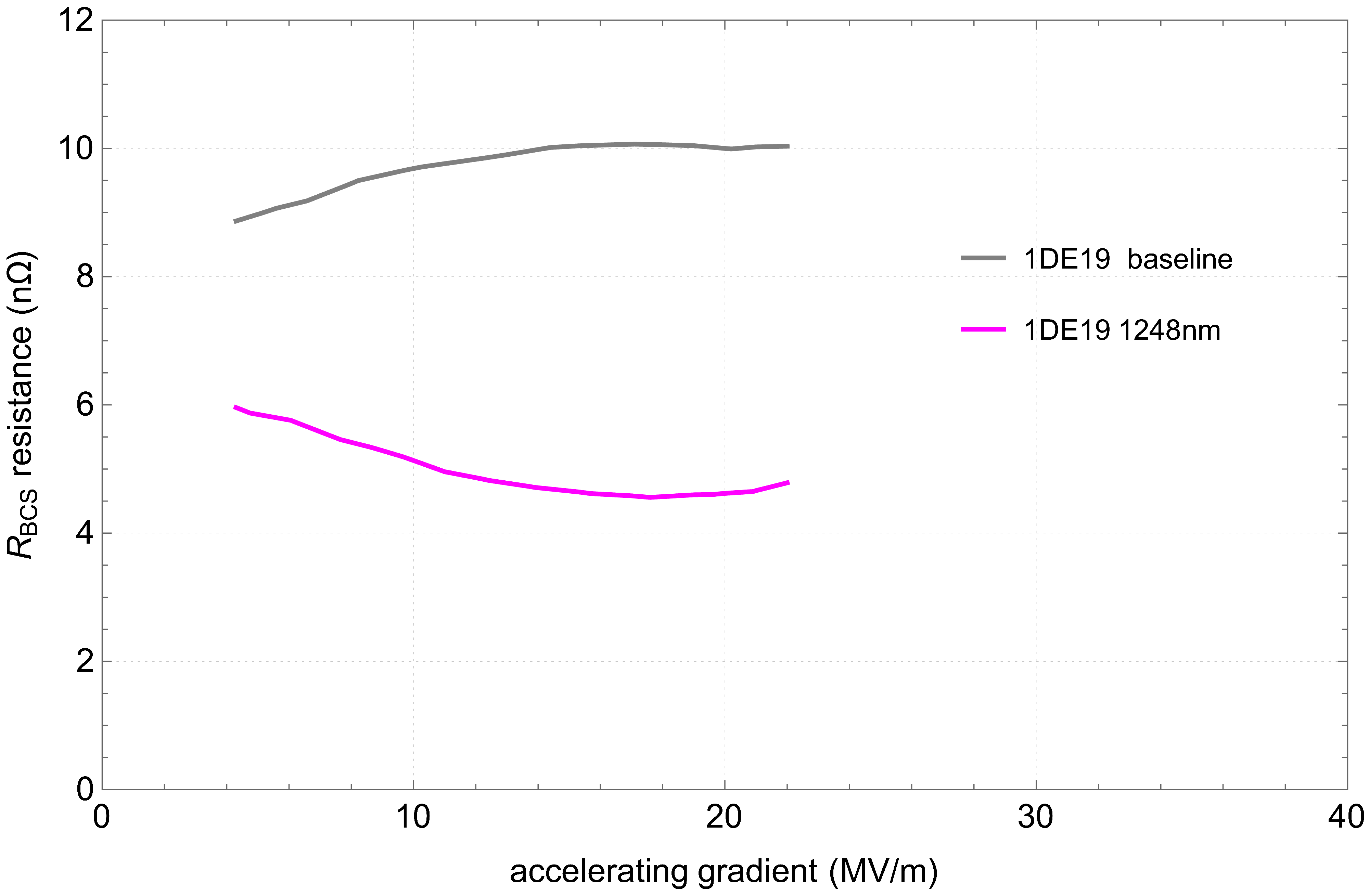}
   \caption{Estimation of $R_{BCS}$ by $R_{BCS,2~\text{K}}\approx R_{S,2~\text{K}}-R_{S,1.5~\text{K}}$ of 1DE19 before the treatment as a baseline and after a mid-T heat treatment of 4.5 h at 335°C corresponding to a calculated oxygen diffusion length of 1248~nm.}
   \label{fig:1DE19_R_BCS}
\end{figure}

The mid-T heating studies are conducted on 1.3~GHz single-cell TESLA-shape cavities with niobium material of RRR~300, manufactured by various suppliers. As baseline treatment, 
cavities undergo at least one extended electropolishing (EP) with the removal of $110-140~\text{µm}$, followed by an 800°C annealing. The final baseline treatment before the mid-T heat treatment consists of a short EP or an 800°C soft-reset, followed by a baseline test. In addition to so-called fine-grain (FG) material, we are investigating this treatment on large-grain (LG) material as well. Additionally, two coated cavities were treated; details on the coatings and results can be found in \cite{Wenskat_2023, doi:10.1021/acs.chemmater.3c03173}.\\
In preparation of later mid-T heat treatments we either start with a short electropolishing (EP) or a 800°C 3-hour so-called 'soft reset'. The soft reset is defined as successful in case the cavity performance (\(Q(E)\)) matches that of a freshly EP'ed cavity. Specifically, we expect a high field Q-slope (HFQS) with resulting power limitation \cite{Reschke_PhysRevAccelBeams.20.042004} at about 30~MV/m, no field emission, and low residual resistance.\\
When evaluating the RF performance of treated cavities, a so-called baseline measurement was always conducted initially. The vertical tests involve capturing \(Q(T)\) as well as \(Q(E)\) curves at 2~K, 1.8~K and 1.5~K where the maximum attainable gradient $E_{acc}$ and its corresponding $Q_0$ are measured. It should be noted, however, that for many of our baseline measurements, only a 2~K curve was recorded. 
%
Especially for mid-T treated cavities \cite{HIto}, the ambient magnetic field significantly influences the $Q_0$ (specifically the RF surface resistance, see below). In addition to the double magnetic shielding \cite{Reschke_PhysRevAccelBeams.20.042004} of the test cryostats, the magnetic hygiene was continuously improved throughout the reported test campaign. Based on earlier measurements, the ambient magnetic field for the vertical tests can be estimated to be $\leq$~0.3~µT. Presently, measurements using a 3-axis magnetometer at 2~K are ongoing in order to gain more precise values. No active field compensation was applied during any of our measurements.
A general overview of the workflow for vertical acceptance tests is shown in \cite{Reschke_PhysRevAccelBeams.20.042004}.\\
%
%
Concerning measurement accuracy, it needs to be stated that the estimated uncertainty for independent RF measurements is approximately $\sim$10\% for $E_{acc}$ and up to $\sim$20\% for $Q_0$. However, within a single cold vertical test
and for each \(Q(E)\) curve, the observed measurement scatter is significantly smaller, around 1\% for $E_{acc}$ and 3\% for $Q_0$ \cite{Reschke_PhysRevAccelBeams.20.042004}. None of the presented cavities were limited by radiation, indicating that we observed no field emission. Instead, their performance was consistently either power-limited or limited by breakdown (quench).\\
We exemplify the result of mid-T heat treatment in Fig.~\ref{fig:1DE19withbaseline} (approximately 4.5~h at 335°C) using cavity 1DE19, where we display one of our first mid-T heating results. Shown is the baseline measurement and the outcome after treatment based on \(Q(E)\) curves at 2~K and 1.5~K. The 4.5-hour treatment at 335°C originated from the commissioning period of the furnace.
%
The surface resistance is given by
\begin{equation}
    R_S(T, B) = R_{BCS}(T) + R_{const.}, 
\end{equation}
where $R_{BCS}(T)$ is the temperature dependent BCS resistance and $R_{const.}=R_{res}+R_{flux}(B)$ consisting of the temperature independent parts, $R_{res}$, the residual resistance and $R_{flux}(B)$, a magnetic flux induced additional surface resistance. Without intentionally applying a magnetic field $R_{flux}$ is dominated by the ambient magnetic environment of the vertical test. Considering that the temperature-dependent contribution to the surface resistance \(R_S\) at 1.5~K is typically \(\leq 1~\text{n}\Omega\) the temperature independent resistance is estimated by $R_{const.}\approx R_{S,1.5~\text{K}}$. Utilizing \(Q(E)\)-curves obtained at 2~K and 1.5~K, we derive a measure that closely approximates \(R_{BCS} \approx R_{S,2~\text{K}} - R_{S,1.5~\text{K}}\) at 2~K. 
This is illustrated in Fig.~\ref{fig:1DE19_R_BCS} for the baseline and after mid-T heat treatment on cavity 1DE19. 
There is a clear reduction in $R_{BCS}$ after the mid-T heat treatment. Additionally, we observe no HFQS and an improvement in the quality factor $Q_0$, albeit with an earlier quench, as shown in Fig.~\ref{fig:1DE19withbaseline}.\\
%
In addition to the cavities, witness samples were also installed during each run. The additional samples are accommodated within recesses on the frame part supporting the upper cavity flange (see Fig. \ref{fig:furnacemodel}). Like the cavities, these samples undergo bulk EP, 800°C annealing and fine EP treatments before the mid-T heat treatment. In a tandem run, two samples are installed at their respective cavity-flange heights.

\subsection{Effective oxygen diffusion length}

\begin{figure}[!htb]
   \centering
   \includegraphics*[width=1.0\columnwidth]{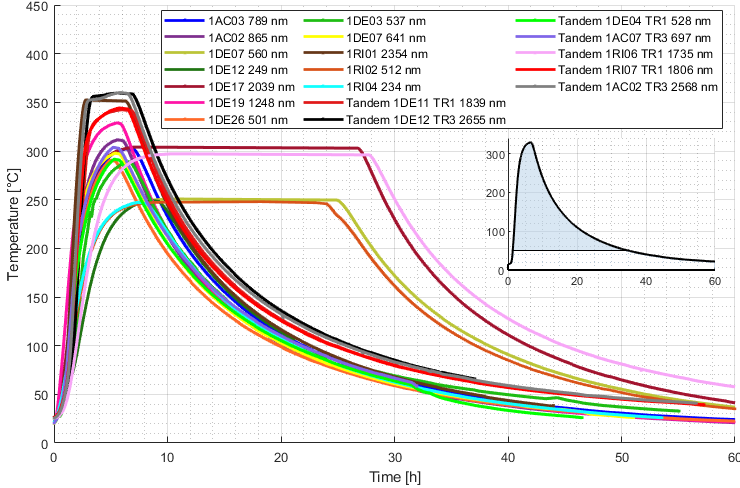}
   \caption{Temperature profiles of all mid-T heat treatments. The corresponding oxygen diffusion length is indicated in the legend. Shown as a small inset on the right is the numerically integrated area of one temperature profile, used to determine the diffusion length.}
   \label{fig:Tprofilesall}
\end{figure}

Activities regarding the significant impact of heating a cavity under vacuum at specific temperatures and durations on \(Q(E)\) performance have long centred on interstitial oxygen. It has been demonstrated that the pentoxide layer on niobium surfaces decomposes within the temperature range of 200°C to 300°C while the monoxide layer remains \cite{palmer1988surface, palmer1990oxide, Kowalski:SRF03-THP09}.\\
In this study presented here, we assume the simplest case, where the only source for oxygen diffusing into the bulk is from the oxide layer. Furthermore, it is assumed that oxygen diffusion into the material from residual gases in the vacuum furnace does not occur during treatment. Accordingly, we employ the simple diffusion model of Fick \cite{FickLaw} and investigate whether the calculated diffusion length $l$, correlates with RF properties. While crystalline solids can exhibit several structurally different paths by which atomic diffusion can take place, we only consider bulk diffusion.
It is known that grain boundaries serve as higher diffusivity paths, and diffusion tends to occur faster along these. Hence initially, it remained unclear whether the mid-T heat treatment would be effective on LG material or if more grain boundaries as in FG material are necessary.
Same treatment duration and temperatures on FG and LG material thus may have different effects on the cavity performance.\\
%
%
The temperature profiles obtained from mid-T heat treatments at DESY, as illustrated in Fig.~\ref{fig:Tprofilesall}, were measured using the temperature sensors depicted in Fig.~\ref{fig:furnacemodel}. 
Integration over the complete temperature profile for temperatures above 50°C is done to calculate the diffusion length,
as contributions are derived from the entire area.\\
The Arrhenius equation
    \begin{equation}
    D(T) = D_0 \cdot e^{-\frac{E_a}{kT(t)}},
    \label{eq:diffusionequation}
    \end{equation}
describes the temperature-dependent diffusion coefficient for oxygen atoms in Nb, where the so-called diffusion activation parameters $D_0 = 0.015~\frac{\text{cm}^2}{\text{s}}$ represents the diffusion constant, $E_a = 1.2~\text{eV}$ denotes the activation energy \cite{DiffusioninSolids}, and $k$ refers to the Boltzmann constant. 
Applying Fick's Law \cite{FickLaw} to equation \ref{eq:diffusionequation}, the diffusion length from the surface can be determined as $l = \sqrt{Dt}$, where $t$ represents time. Considering the entire heat cycle of the furnace run introduces a time-dependent diffusion coefficient due to varying temperatures. To address this, numerical integration of the diffusion coefficient over time is performed: $l=\sqrt{\int D(t) dt}$, utilizing data from the respective temperature sensor. 
Figure \ref{fig:Tprofilesall} displays temperature profiles over time for each mid-T heat treatment, along with corresponding oxygen diffusion lengths indicated in the legend and in Table \ref{tab:thermalbudget}. 

\section{Mid-T campaign}

Since the initiation of the campaign, a substantial amount of data has been gathered, covering a total of 19 treatments across 17 cavities. Mid-T recipes in the temperature range 250-350°C and holding times from 3-20~h were tested. 

\subsection{$Q_0$ versus $E_{acc}$}

\begin{figure*}[!htb]
    \centering
        \subfigure[]{\includegraphics*[width=1\columnwidth]{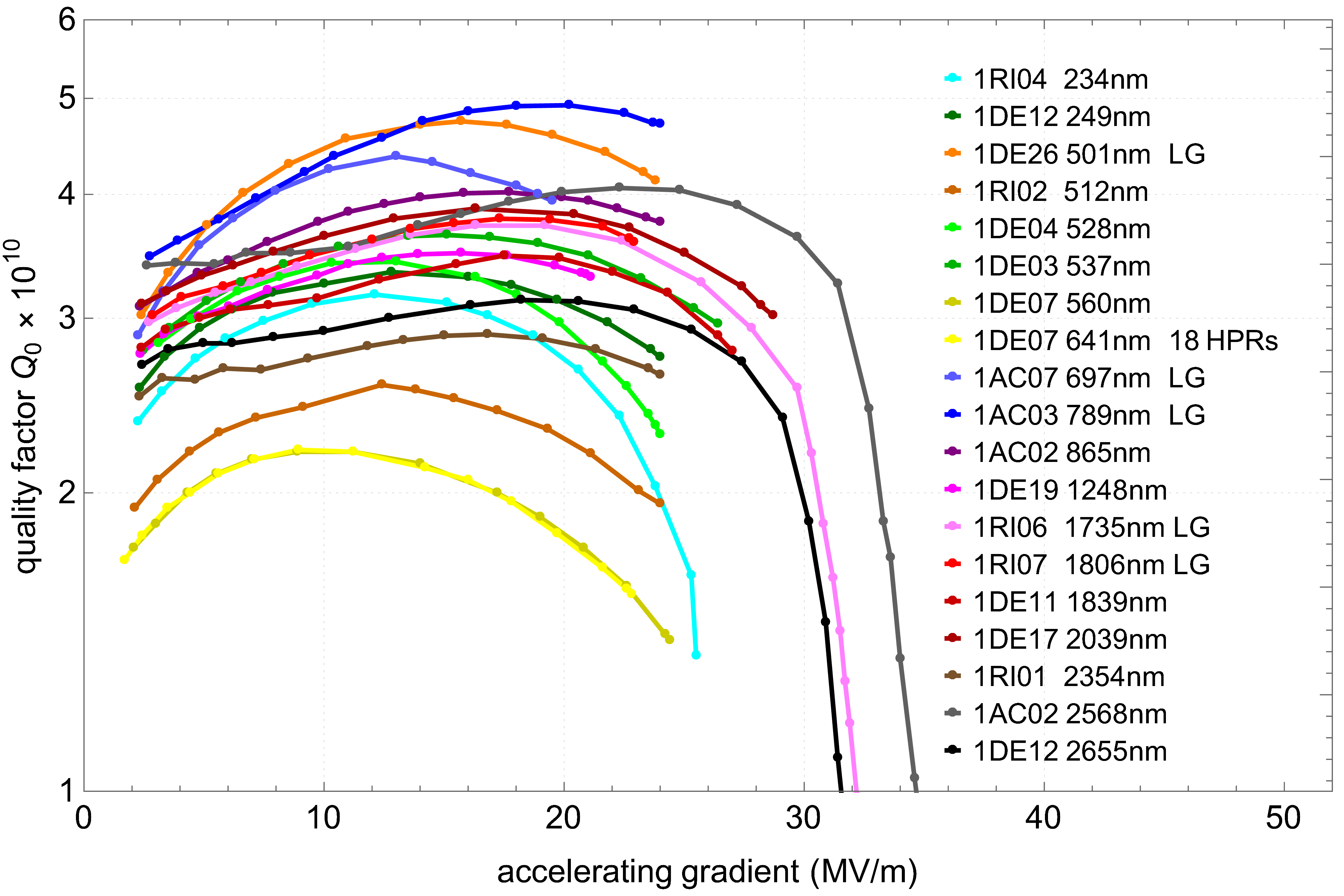}}
        \subfigure[]{\includegraphics*[width=1\columnwidth]{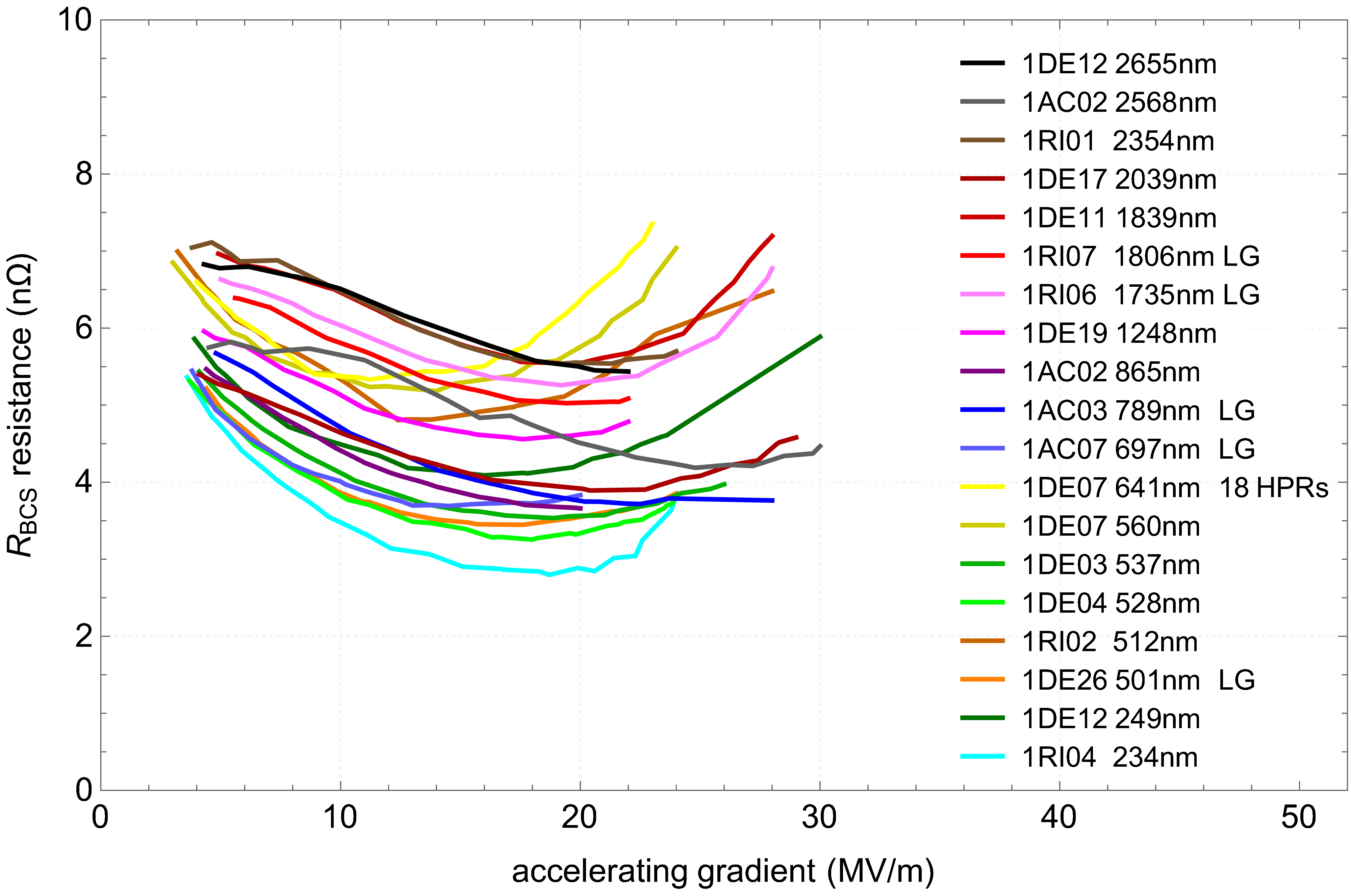}}
    \caption{\(Q(E)\) and $R_{BCS}$ at 2~K for all mid-T heat treatments. The calculated oxygen diffusion length of each treatment is indicated. \(Q(E)\) data representing the final treatment status of the respective vertical tests, including the gradient achieved (in some instances the test was intentionally terminated at 24~MV/m before quenching to investigate the maximum $Q_0$) are shown in (a). Estimation of $R_{BCS}$ by $R_{BCS,2~\text{K}} \approx R_{S,2~\text{K}}-R_{S,1.5~\text{K}}$ are shown in (b).}
     \label{fig:QvsEall_Rbcs_all}
\end{figure*}

The graphical representation of $Q_0$ at 2~K versus $E_{acc}$ is presented in Fig.~\ref{fig:QvsEall_Rbcs_all} (a), while Fig.~\ref{fig:QvsEall_Rbcs_all} (b) illustrates the approximated $R_{BCS}$ at 2~K.
Distinctive features of an anti-Q slope, with a maximum in $Q_0$ in the range of 16~MV/m to 20~MV/m of mid-T treated cavities are evident in Fig.~\ref{fig:QvsEall_Rbcs_all}~(a). The accelerating gradient limit is often below 30~MV/m.
While many similarities are observed, the values notably differ, particularly in terms of quality factors, which will be further discussed in the subsequent sections. It has been addressed for example in \cite{Jacek_CW_16MVm} that cavities with accelerating gradients of approximately 16~MV/m are required for a CW upgrade for the European XFEL. In addition, current SRF applications such as the LCLS-II demonstrate the need for gradients of 16~MV/m \cite{galayda2018lcls}.

\subsection{$Q_0$ degradation}

\begin{figure}[!htb]
   \centering
   \includegraphics*[width=1.0\columnwidth]{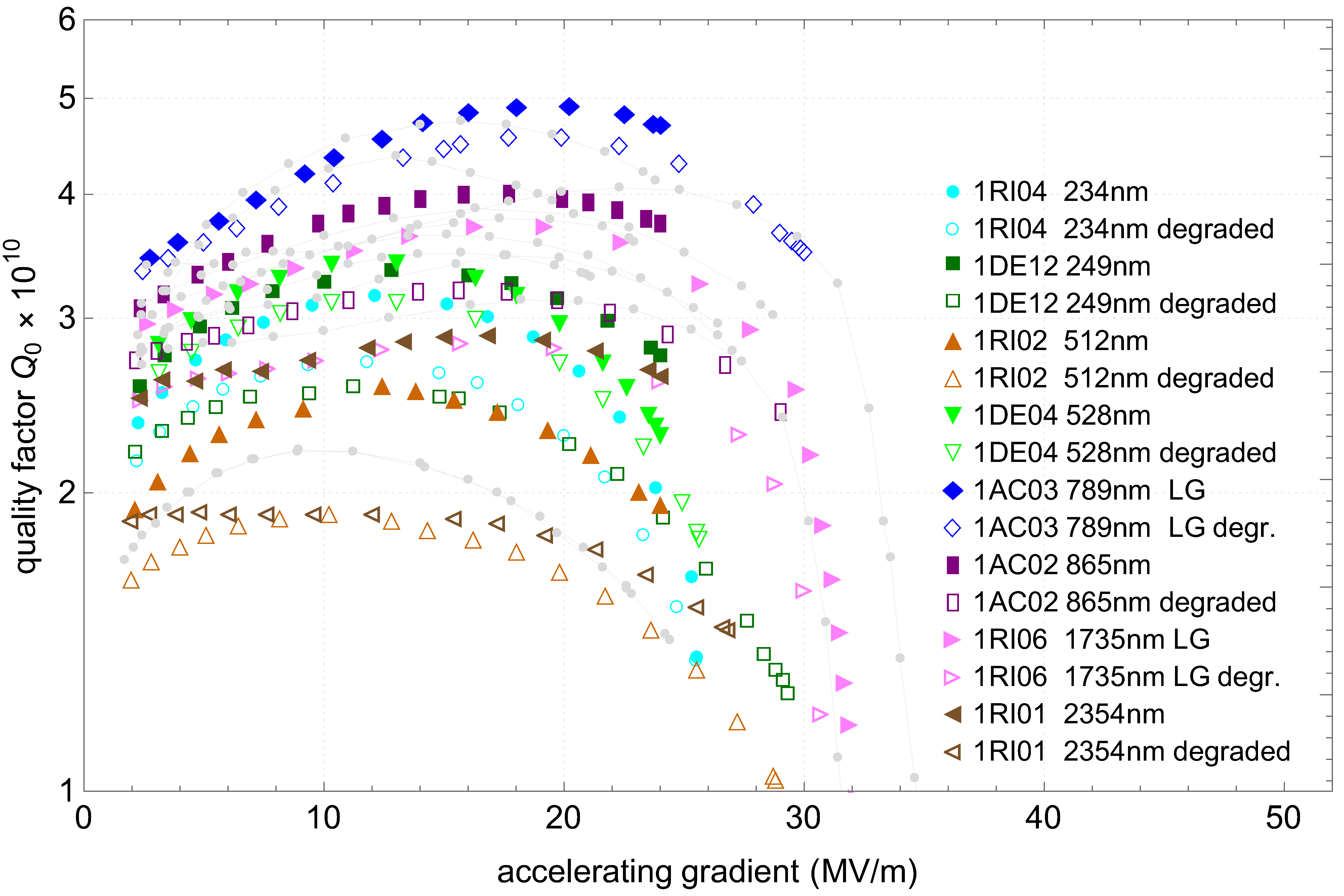}
   \caption{\(Q(E)\) at 2~K curves for all mid-T heat treatments. The grey data points denote cavity performance unaffected by $Q_0$ degradation after quenching. Filled coloured symbols indicate curves intentionally terminated before quenching to investigate the maximum $Q_0$. Empty symbols represent data indicating $Q_0$ degradation after quenching and the gradient reach.}
   \label{fig:QvsE_degraded}
\end{figure}

Among the presented measurements, eight show a decrease in $Q_0$ after quenching as shown in Fig.~\ref{fig:QvsE_degraded}. No systematic cause for the observed degradation has been identified thus far. However, the quality factor can always be restored when the temperature is warmed up to about 30~K and cooled down again below $T_c$ which is referred to as thermal cycling. The degradation does not appear to be related to the oxygen diffusion length, the use of FG or LG material, or the manufacturer. Additionally, instances occurred where a cavity did not exhibit this effect after the first treatment but did so after a second treatment. This suggests that the cause may lie within the treatment process itself. However, it is crucial to understand the cause of this effect and to find a way to mitigate it in case mid-T heat treated cavities are used in accelerator modules.

\subsection{Estimated $R_{BCS}$}

$R_{BCS}$ for all treatments are shown in Fig.~\ref{fig:QvsEall_Rbcs_all}~(b). For gradients at the study point of around 16~MV/m, the estimated $R_{BCS}$ ranges from 3 to 6~n$\Omega$ after mid-T heat treatments with temperatures below 350°C. Moreover, a distinctive concave curvature is evident, leading to the characteristic anti-Q-slope.\\
The $R_{BCS}$ for the baseline test following electropolishing or a soft-reset is with 9-12 n$\Omega$ significantly higher and the downward curvature cannot be observed as shown exemplarily in Fig. \ref{fig:1DE19_R_BCS}.

\subsection{Mid-T heat parameters sorted by effective diffusion length}

Tab. \ref{tab:thermalbudget}, shows a list of all treated cavities organized according to their calculated effective diffusion length.
Due to the not trivial temperature control (see Fig. \ref{fig:overshoot}) of the furnace the temperature profiles are varying as shown in Fig. \ref{fig:Tprofilesall}.
%
The optimization process of the T-cycles resulted in a variation of the real $T(t)$-curves and thus in a variation of the effective diffusion lengths.
The information for the recipe labeling in Tab. \ref{tab:thermalbudget} is nominal. 
%
It is therefore important to note that, owing to the profile variations, the 3-hour 300°C treatment does not consistently yield the same effective diffusion length across all cases.\\
To systematically analyze and categorize the diverse ranges of treatments conducted, we have opted to group them according to their diffusion length. Large-grain cavities are included within the respective groupings but will be examined in more detail separately in the subsequent section.

\begin{table}[]
    \centering
    \begin{tabular}{c|c|c|c}
    Cavity & Nominal & $l$ (nm)  & \\
     & treatm. &  & \\  
    \hline
    1RI04    & <3h <250°C  &  234 & Fig. 9 \\ 
    1DE12    & <3h 250°C  &  249 & short $l$ \\ 
    \hline
    1DE26 (LG)    & <3h <300°C  &  501 &  \\ 
    1RI02    & 20h 250°C  &  512 & \\
    1DE04    & <3h <300°C  &  528 & \\ 
    1DE03    & <3h <300°C  &  537 & Fig. 10 \\
    1DE07    & 20h 250°C  &  560 & medium $l$\\
    1DE07 18xHPR    & <3h <300°C  &  641 & \\
    1AC07 (LG)   & <3h <300°C  &  697 & \\
    1AC03 (LG)   & <3h <300°C  &  789 & \\
    1AC02    & 3.25h <325°C  &  865 & \\
    \hline
    1DE19    & 4.5h <335°C  &  1248 & Fig. 4 \\
    \hline
    1RI06 (LG)   & 20h 300°C &  1735 &  \\ 
    1RI07 (LG) & 3h <350°C & 1806 & \\
    1DE11    & 3h <350°C  &  1839 & Fig. 11\\
    1DE17    & 20h 300°C  &  2039 & long $l$\\
    1RI01    & 3h 350°C  &  2354 & \\
    1AC02    & 3h $\geq$350°C  &  2568 & \\
    1DE12    & 3h >350°C  &  2655 & \\
    \end{tabular}
    \caption{Cavity treatments sorted by calculated diffusion length $l$ and grouped according to $l$.}
    \label{tab:thermalbudget}
\end{table}
    

\section{Results of the individual effective diffusion lengths}

\subsection{Treatments with short diffusion lengths below that of 3h 300°C}
\begin{figure*}[!htb]
    \centering
        \subfigure[]{\includegraphics*[width=1\columnwidth]{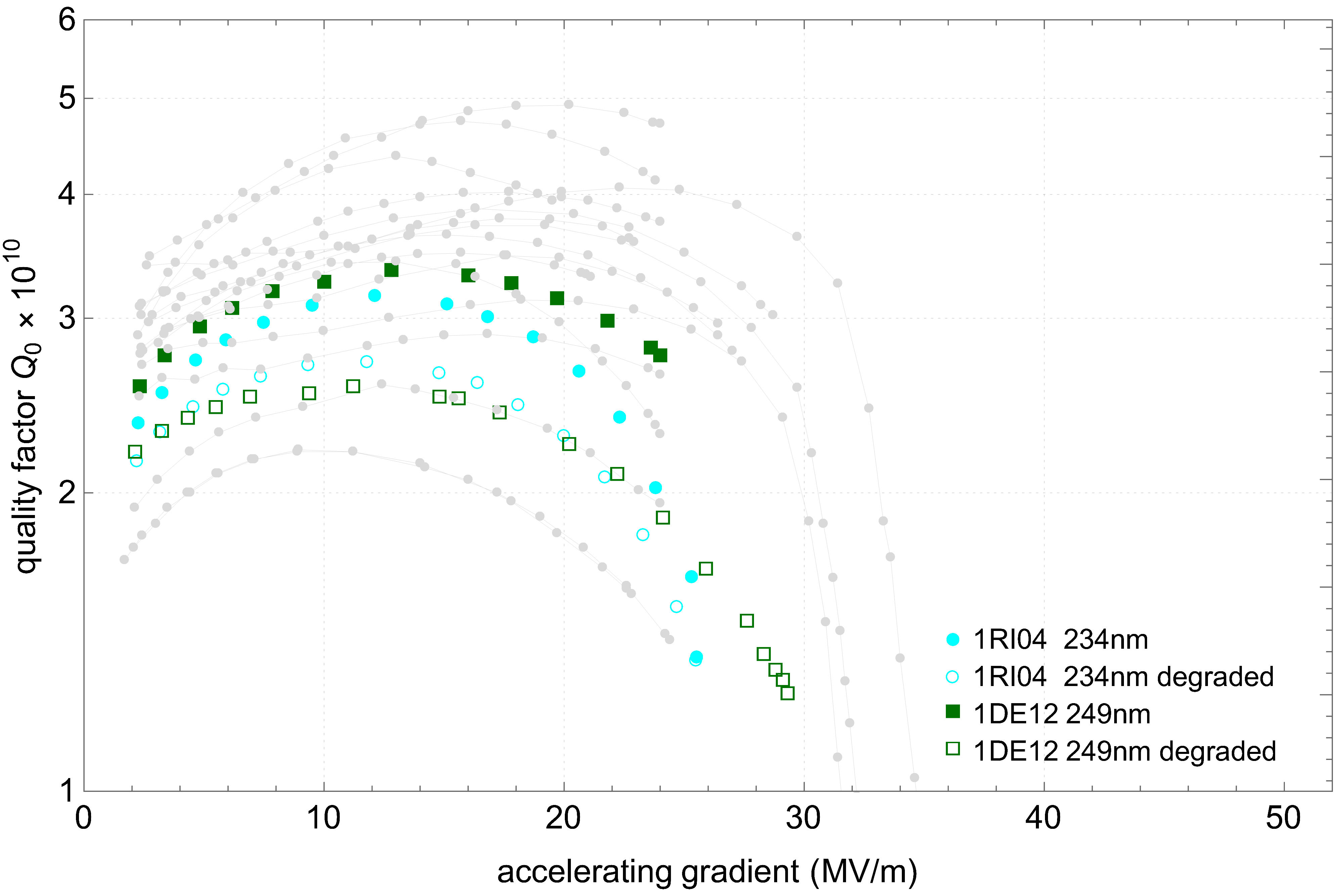}}
        \subfigure[]{\includegraphics*[width=1\columnwidth]{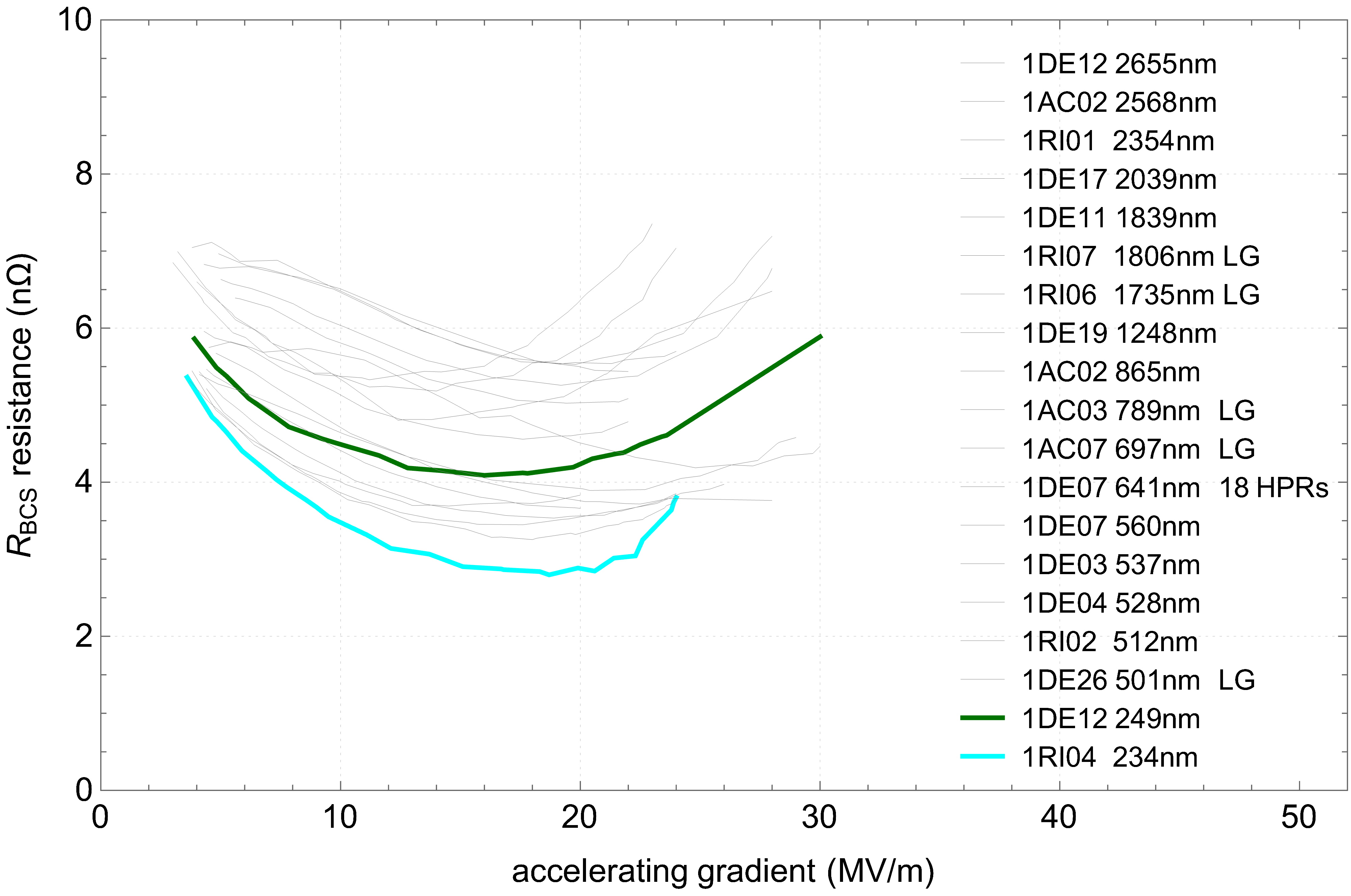}}
    \caption{\(Q(E)\) and $R_{BCS}$ at 2~K for cavities subjected to an oxygen diffusion length closely resembling that of a 3-hour treatment at 250°C. \(Q(E)\) data representing the final treatment status of the respective vertical tests, including the gradient achieved (in one instance the test was intentionally terminated at 24~MV/m before quenching to investigate the maximum $Q_0$) are shown in (a). Estimation of $R_{BCS}$ by $R_{BCS,2~\text{K}} \approx R_{S,2~\text{K}}-R_{S,1.5~\text{K}}$ are shown in (b).}
     \label{fig:QvsE_250_RBCS_250}
\end{figure*}

Here, we consider diffusion lengths between 234 and 249~nm, corresponding to treatments similar to 3 hours at 250°C. Refer to the first group in Table \ref{tab:thermalbudget} where two cavities were treated. The $Q_0$ vs $E_{acc}$ outcomes
are depicted in Fig.~\ref{fig:QvsE_250_RBCS_250}~(a), while the corresponding estimated $R_{BCS}$ are illustrated in Fig.~\ref{fig:QvsE_250_RBCS_250}~(b). Quality factors ranging from 2.5 to $3\cdot10^{10}$ were achieved. Before the degradation, gradients of no more than 25~MV/m at $Q_0>2 \cdot 10^{10}$ were achieved. 
One of the cavities showcases the lowest BCS resistance among all investigated cases. Notably, in all instances, the $Q_0$ degrades after quenching.\\

\subsection{Treatments with medium diffusion lengths similar to that of 3h 300°C}
\begin{figure*}[!htb]
    \centering
        \subfigure[]{\includegraphics*[width=1\columnwidth]{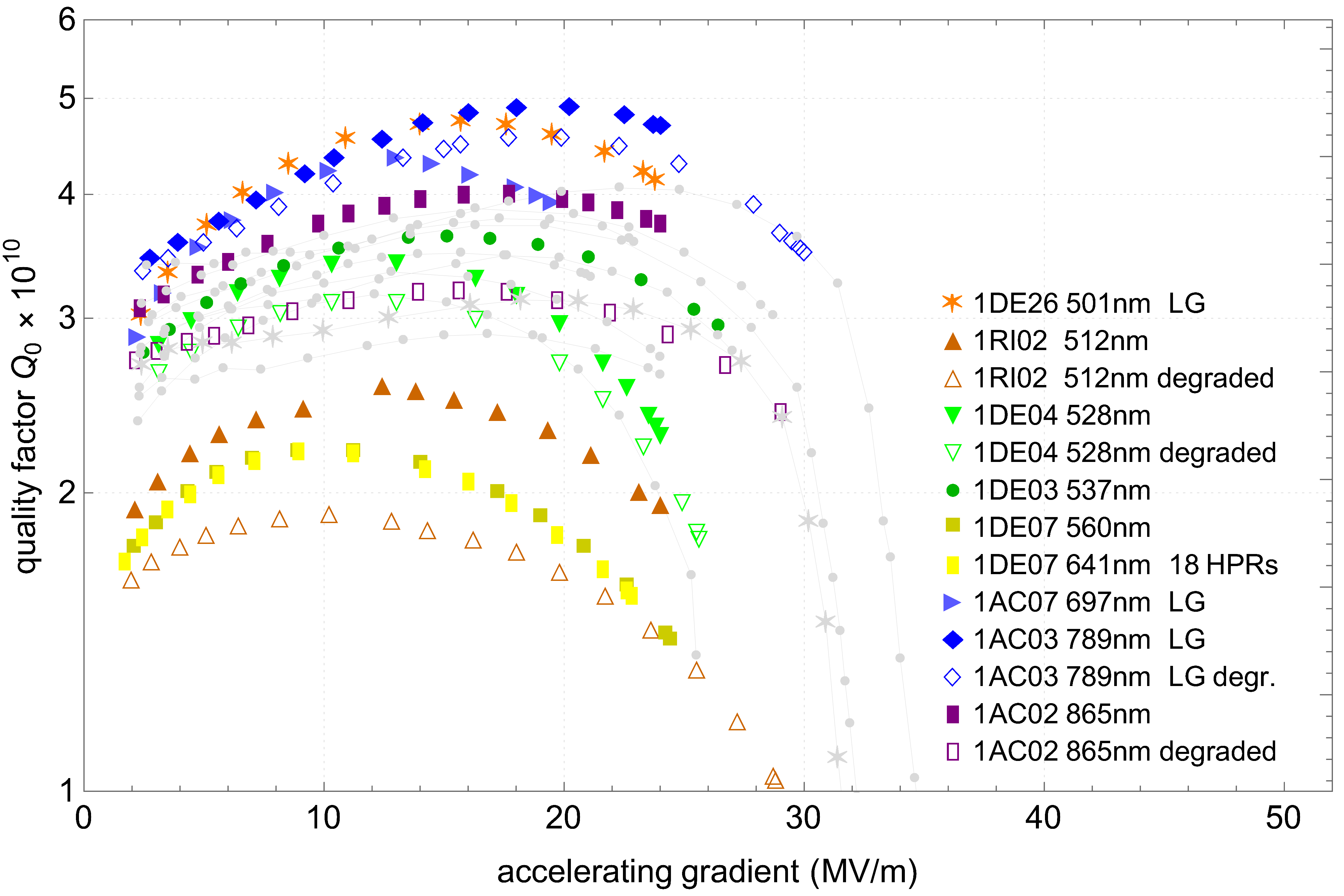}}
        \subfigure[]{\includegraphics*[width=1\columnwidth]{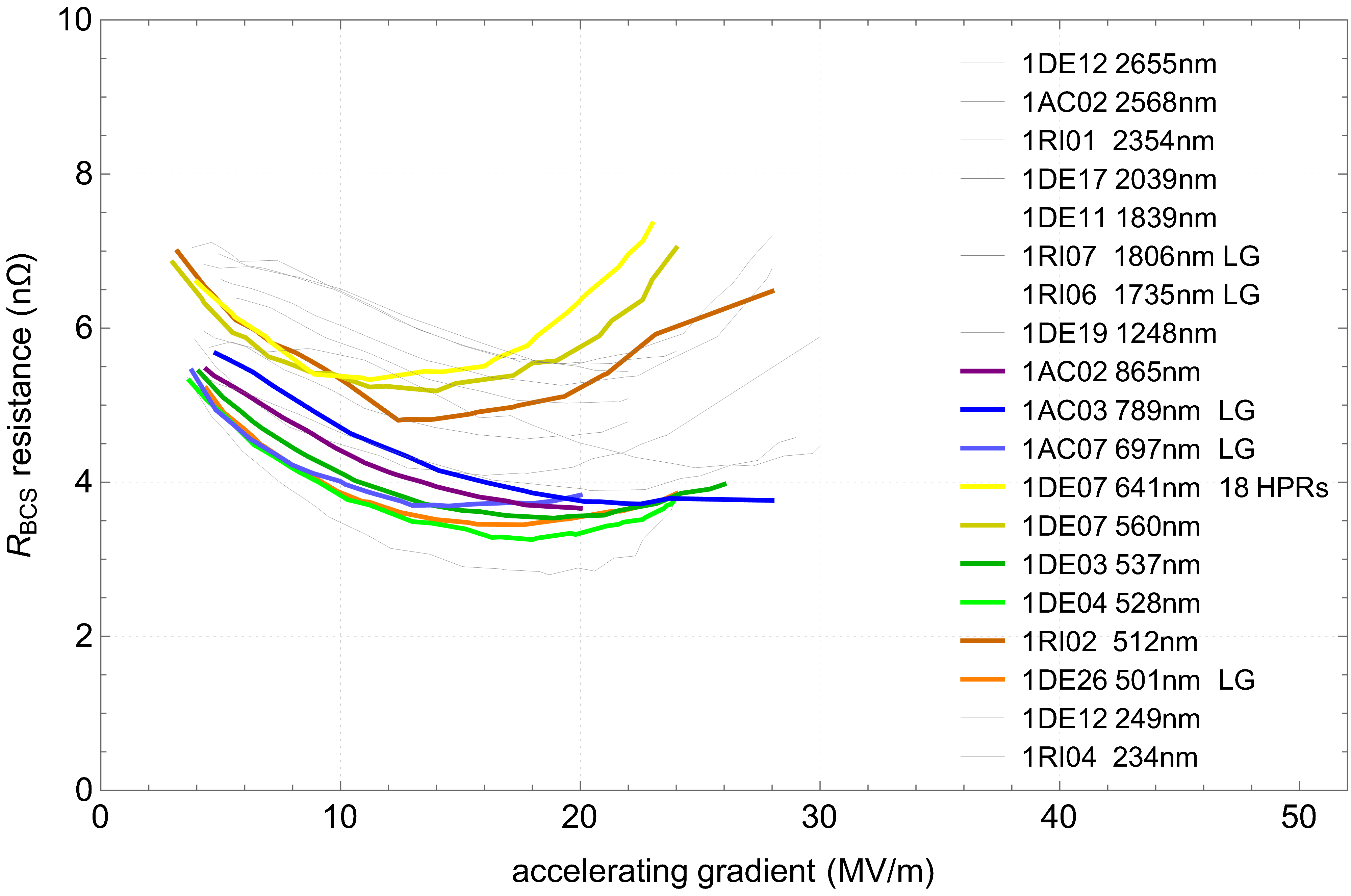}}
    \caption{\(Q(E)\) and $R_{BCS}$ at 2~K for cavities subjected to an oxygen diffusion length closely resembling that of a 3-hour treatment at 300°C. The cavity (1DE07) with the extended x18 HPR and the two cavities (1DE07 and 1RI02) subjected to the prolonged 20-hour 250°C treatment stand out distinctly from the rest. \(Q(E)\) data representing the final treatment status of the respective vertical tests, including the gradient achieved (in all not-degraded instances the test was intentionally terminated at 24~MV/m before quenching to investigate the maximum $Q_0$) are shown in (a). Estimation of $R_{BCS}$ by $R_{BCS,2~\text{K}} \approx R_{S,2~\text{K}}-R_{S,1.5~\text{K}}$ are shown in (b).}
     \label{fig:QvsE_300_RBCS_300}
\end{figure*}

A total of nine cavities were treated, with diffusion lengths ranging from 501 to 1248~nm, as listed in Group 2 of Table~\ref{tab:thermalbudget}, corresponding to the 3-hour 300°C treatment. A 300°C treatment consistently exhibits remarkably high $Q_0$ up to $4\cdot10^{10}$ for FG and $5\cdot10^{10}$ for LG material with a distinct anti-Q-slope, illustrated in Fig.~\ref{fig:QvsE_300_RBCS_300}~(a). In Fig.~\ref{fig:QvsE_300_RBCS_300}~(b) the favourable $R_{BCS}$ are demonstrated. Four out of nine cavities in this group show degradation. Gradients between 25 to 28~MV/m were achieved.\\
However, three specific cases deviate from this trend: Two 20-hour 250°C treatments (1RI02 512~nm and 1DE07 560~nm), despite their prolonged duration and the resulting comparable diffusion length, 250°C appears to produce lower $Q_0$.
Additional 18 high-pressure water rinsings (HPRs) were applied on 1DE07 to study the influence on oxide thicknesses \cite{Rezvan_Marc_talk_TTC_23_US}.
While sharing a similar diffusion length (641~nm), the cavity exhibits a significantly different $R_{BCS}$ as can be well observed in Fig.~\ref{fig:QvsE_300_RBCS_300}~(b), triggering further investigation into the influence of prior-treatment high-pressure water rinsings.

\subsection{Treatments yielding longer diffusion lengths than that of 3h 300°C}

\begin{figure*}[!htb]
    \centering
        \subfigure[]{\includegraphics*[width=1\columnwidth]{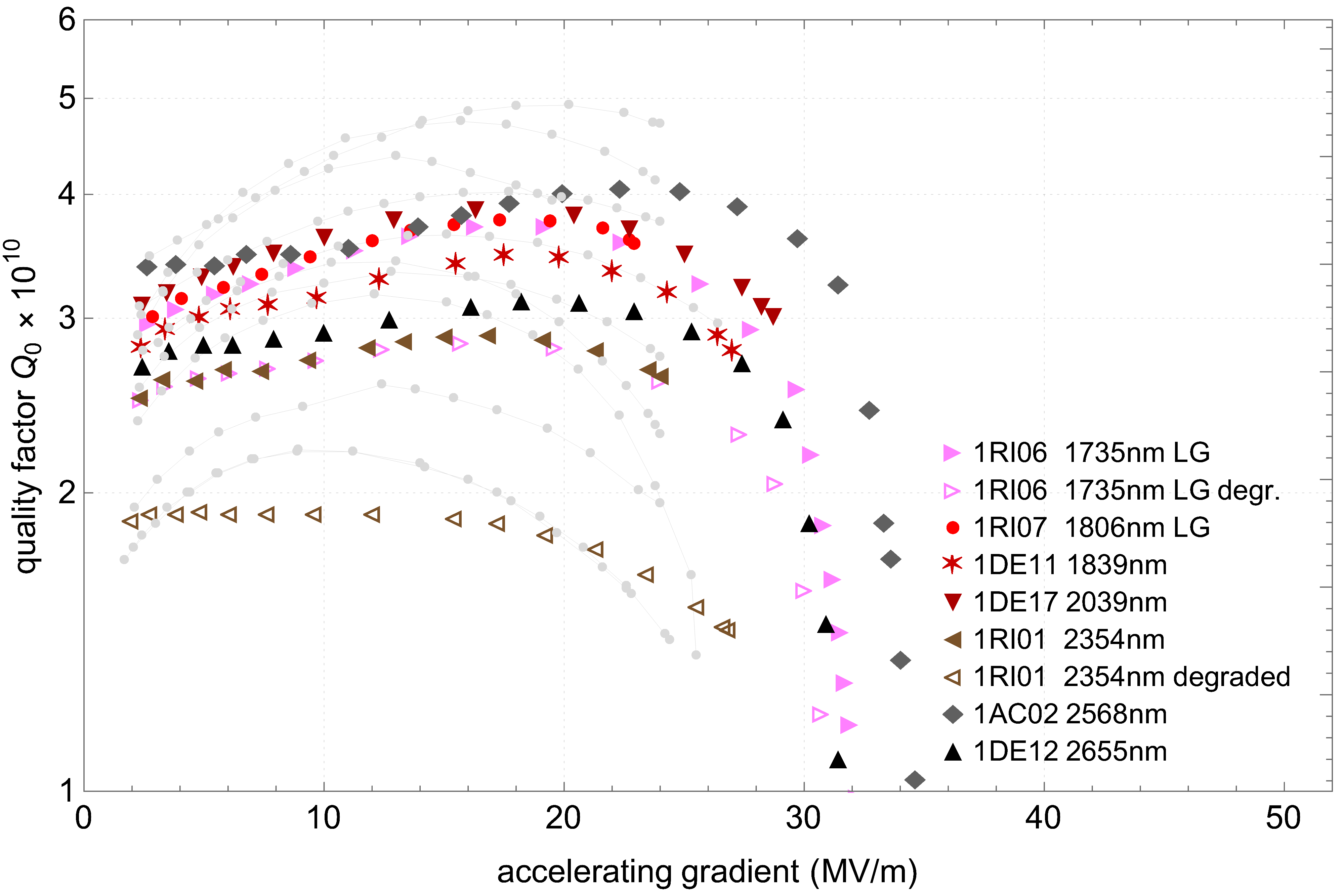}}
        \subfigure[]{\includegraphics*[width=1\columnwidth]{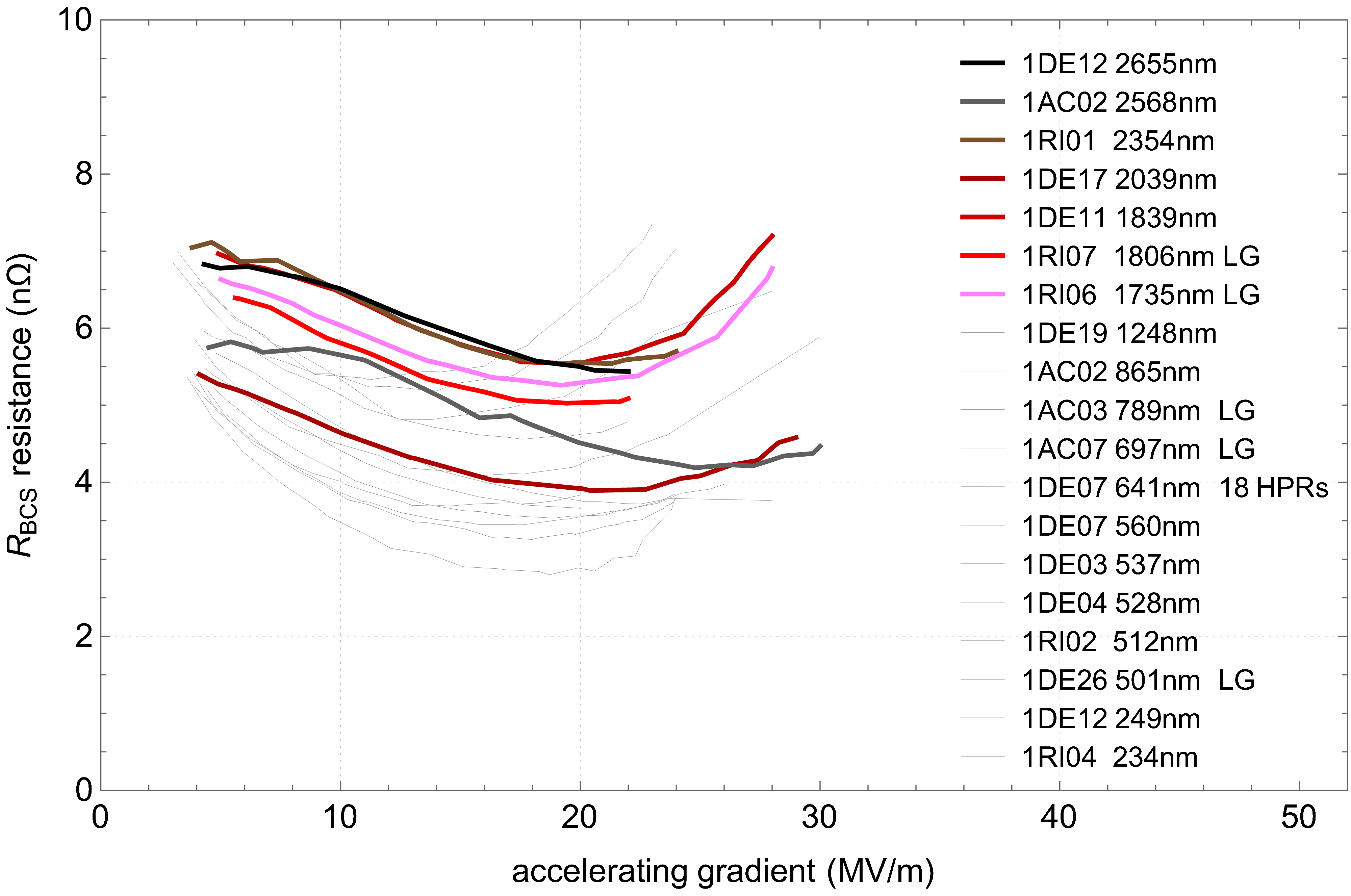}}
    \caption{\(Q(E)\) and $R_{BCS}$ at 2~K for cavities subjected to an oxygen diffusion length closely resembling that of a 3-hour treatment at 350°C. \(Q(E)\) data representing the final treatment status of the respective vertical tests, including the gradient achieved (in one instance the test was intentionally terminated at 24~MV/m before quenching to investigate the maximum $Q_0$) are shown in (a). Estimation of $R_{BCS}$ by $R_{BCS,2~\text{K}} \approx R_{S,2~\text{K}}-R_{S,1.5~\text{K}}$ are shown in (b).}
     \label{fig:QvsE_350_RBCS_350}
\end{figure*}

A total of seven cavities were treated, with diffusion lengths ranging from 1735 to 2655~nm, as listed in Group 2 of Table \ref{tab:thermalbudget}. In the case of the 350°C treatment, observations reveal a similar $Q_0$ compared to medium $l$ with a less pronounced anti-Q slope, 
as depicted in Fig.~\ref{fig:QvsE_350_RBCS_350}~(a). Notably, the finding emerges with a pronounced high-field Q-slope (HFQS) at 350°C also visible for $R_{BCS}$ as illustrated in Fig.~\ref{fig:QvsE_350_RBCS_350}~(b). The $R_{BCS}$ resistance exhibits significantly different behaviour, with the minimum shifting towards higher gradients. Gradients exceeding 30~MV/m can be achieved but with significantly lower quality factors. The well-known phenomenon HFQS was not expected to occur after a heat treatment at a low temperature of 350°C. The HFQS occurrence may be attributed to the absence of pentoxides at the cavity surface, underscoring the critical role of temperature in the elimination of specific oxides.\\
%

\section{Treatments of large-grain cavities}

\begin{figure*}[!htb]
    \centering
        \subfigure[]{\includegraphics*[width=1\columnwidth]{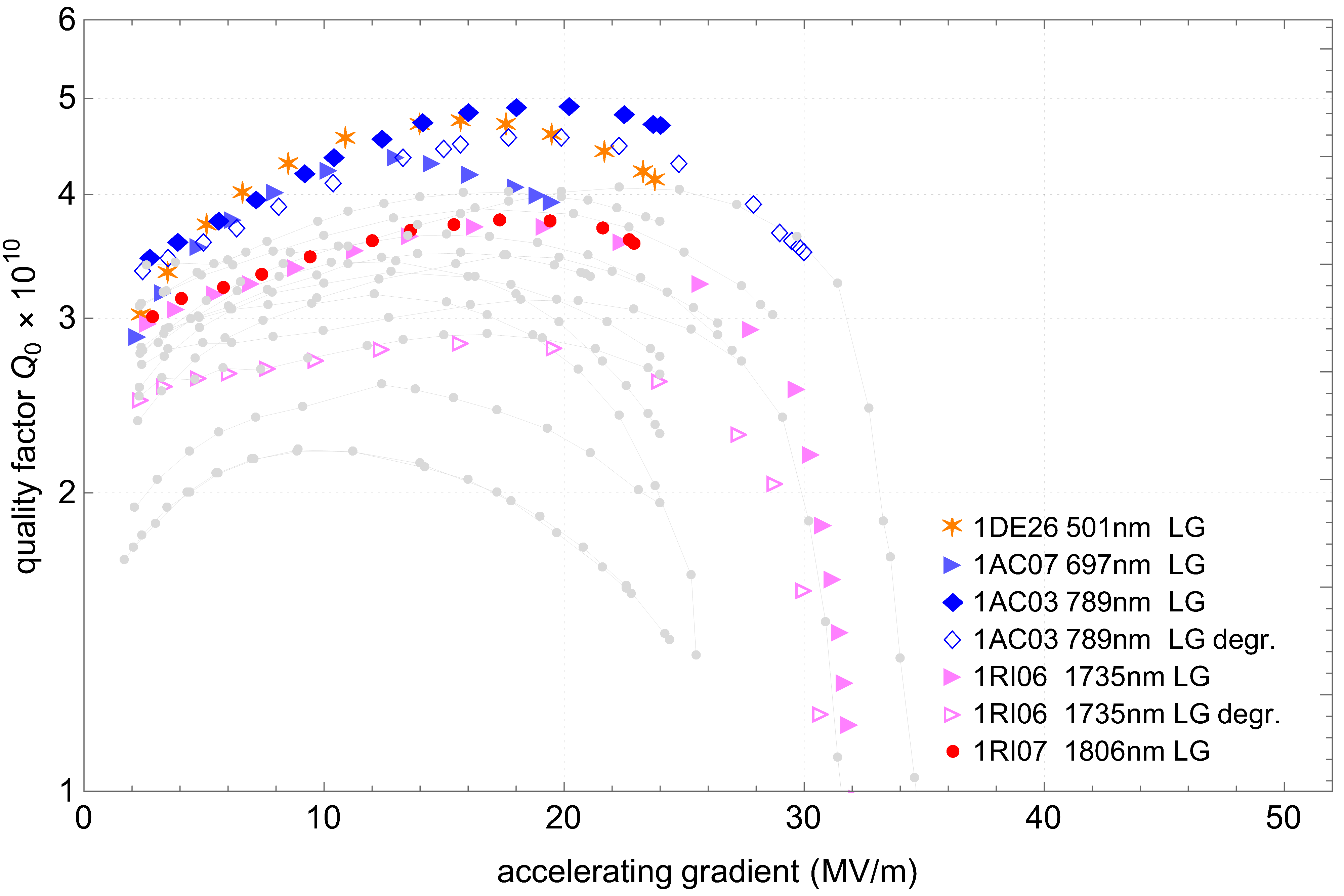}}
        \subfigure[]{\includegraphics*[width=1\columnwidth]{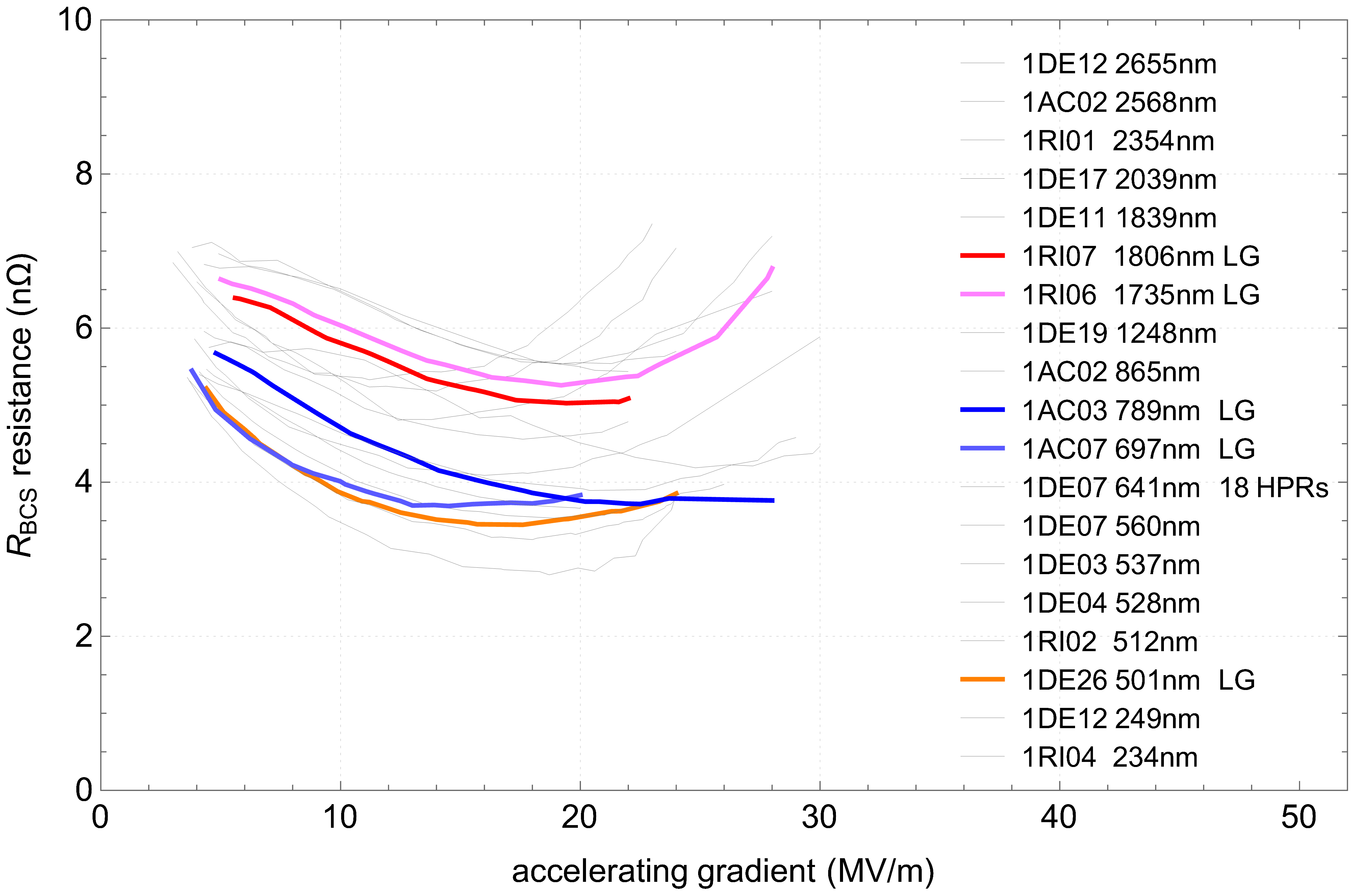}}
    \caption{\(Q(E)\) and $R_{BCS}$ at 2~K for large-grain cavities subjected to a mid-T heat treatment. \(Q(E)\) data representing the final treatment status of the respective vertical tests, including the gradient achieved (in some instances the test was intentionally terminated at 24~MV/m before quenching to investigate the maximum $Q_0$) are shown in (a). Estimation of $R_{BCS}$ by $R_{BCS,2~\text{K}} \approx R_{S,2~\text{K}}-R_{S,1.5~\text{K}}$ are shown in (b).}
     \label{fig:QvsE_LG_RBCS_LG}
\end{figure*}

A total of five large-grain cavities were treated, with diffusion lengths ranging from 501 to 1806 nm. Fig.~\ref{fig:QvsE_LG_RBCS_LG}~(a) presents the outcome of mid-T heat treatments applied to LG single-cell cavities, revealing a noteworthy performance in high $Q_0$. The $R_{BCS}$ values observed in LG cavities shown in Fig. \ref{fig:QvsE_LG_RBCS_LG}~(b) exhibit similarities to those of FG cavities, highlighting comparable behaviour. Notably, for LG cavities, the quality factors are better for diffusion lengths below 1000~nm compared to those above 1000~nm (compare Fig.~\ref{fig:QvsE_LG_RBCS_LG}~(a) and \ref{fig:Q0vsdifflength}), which is also directly reflected in the $R_{BCS}$ resistance (see Fig.~\ref{fig:QvsE_LG_RBCS_LG}~(b)). Additionally, the HFQS also appears in large-grain cavities after a 350°C treatment.\\
The primary improvement in LG cavities compared to FG material in general is attributed to a reduction of temperature independent surface resistance ($R_{const.}$). However, this is a known behaviour of the material \cite{Schlander:154182}, which can not be associated with the mid-T heat treatment.\\
%
Although the diffusion process is known to be enhanced at grain boundaries in polycrystalline materials \cite{DiffusioninSolids}, the contributions of these boundaries seem unnecessary since the mid-T heat treatment is also successfull on LG cavities.
%
%
%

\section{Correlation between cavity performance and effective diffusion length}

\begin{figure}[!htb]
   \centering
   \includegraphics*[width=1.0\columnwidth]{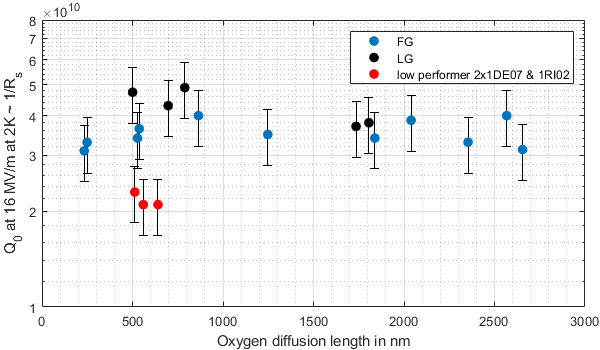}
   \caption{$Q_0$ at 2~K at 16~MV/m value for all mid-T heat treatments against oxygen diffusion length.}
   \label{fig:Q0vsdifflength}
\end{figure}

\begin{figure}[!htb]
   \centering
   \includegraphics*[width=1.0\columnwidth]{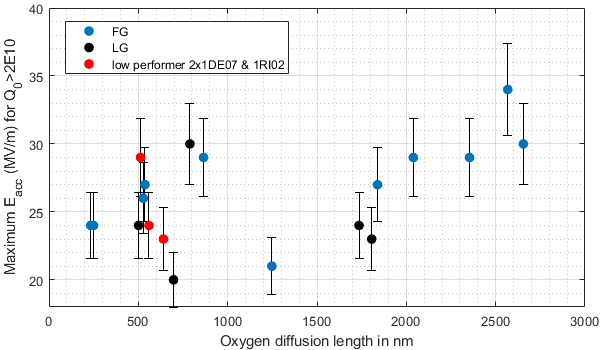}
   \caption{Maximal achieved gradient for $Q_0>2\cdot10^{10}$ for all mid-T heat treatments against oxygen diffusion length.} 
   \label{fig:Emaxvsdifflength}
\end{figure}

\begin{figure}[!htb]
   \centering
   \includegraphics*[width=1.0\columnwidth]{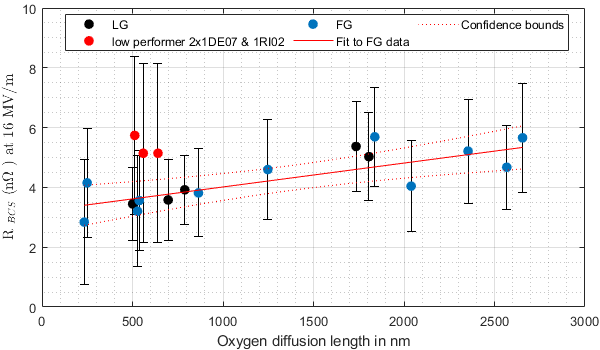}
   \caption{$R_{BCS,2~\text{K}} \approx R_{S,2~\text{K}}-R_{S,1.5~\text{K}}$ at 16~MV/m for all mid-T heat treatments against diffusion length. Large-grain, prolonged HPR treated and 20~h 250°C treated cavities are highlighted. A nonlinear least squares fit of the FG data is depicted in red, with the 95\% confidence bounds represented by the dotted lines.}
   \label{fig:RBCS_vsdifflength}
\end{figure}

\begin{figure}[!htb]
   \centering
   \includegraphics*[width=1.0\columnwidth]{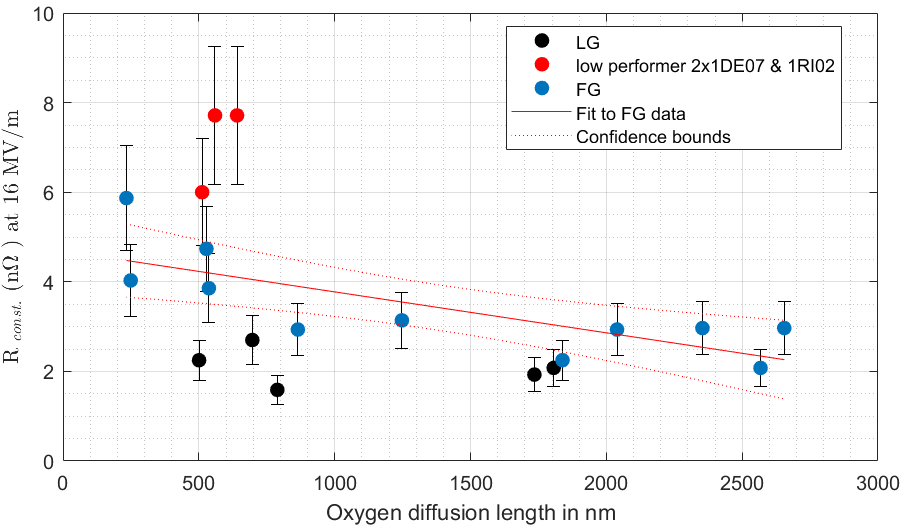}
   \caption{Surface resistance at 1.5~K, nearly equivalent to the temperature independent resistance $R_{const.}$ for all mid-T heat treatments against diffusion length. Values are recorded at 16~MV/m field strength. Large-grain, prolonged HPR treated and 20h 250°C treated cavities are highlighted. A nonlinear least squares fit of the FG data is depicted in red, with the 95\% confidence bounds represented by the dotted lines.}
   \label{fig:Rres_vsdifflength}
\end{figure}

To establish a fixed value for further investigations, 16~MV/m was selected, as this is a foreseen operational gradient for accelerator modules in CW mode \cite{Jacek_CW_16MVm}. Additionally, the maximum quality factor for the 3-hour 300°C treatment recipe occurs around this value. The achieved $Q_0$ values at 16~MV/m are compared to the calculated oxygen diffusion length in Fig.~\ref{fig:Q0vsdifflength}. It is noteworthy to mention that almost all of the treatments shown here achieve a maximum $Q_0$ exceeding $3\cdot10^{10}$. Given the limited precision of the data, identifying a clear trend is difficult.\\
%
%
Examining the maximum achieved accelerating gradient relative to the diffusion length, as illustrated in Fig.~\ref{fig:Emaxvsdifflength}, we notice a possible trend for an increase in gradients for higher diffusion lengths. However, to confirm this statement more data is needed between diffusion lengths of 1000 to 1500~nm. Gradients above 35~MV/m seem not to be in reach for $Q_0>2\cdot10^{10}$ in the parameter range investigated here, which is in agreement with other studies \cite{He_2021,HIto}. Therefore, it is our concern to investigate even higher diffusion lengths further.\\ 
The BCS resistance against the oxygen diffusion length is plotted in Fig.~\ref{fig:RBCS_vsdifflength}, and the temperature independent resistance against the diffusion length is illustrated in Fig.~\ref{fig:Rres_vsdifflength}. 
The cavities that performed slightly worse (extra-long HPR and 20-hour 250°C treatment, see legend in Fig.~\ref{fig:Q0vsdifflength}) are excluded in the regression analysis since these are exceptions as can be seen in particular from the noticeably higher BCS resistances in Fig. \ref{fig:QvsE_300_RBCS_300} (b). In addition, LG cavities were also not included in the regression.
%
By regression analysis, trends become apparent. The BCS resistance seems to be following an upward trend and the residual resistance a downward trend. Thus, we should expect that an optimum in $Q_0$ for diffusion lengths must exist, since $Q_0\sim1/R_S$ with $R_S=R_{BCS}+R_{const.}$. However, because these effects are only weakly pronounced, they cancel each other out, resulting in no discernible optimum in our data, as seen in Fig.~\ref{fig:Q0vsdifflength}.\\
%
%
%
%
Furthermore, it is evident that the improvement in the performance of LG material compared to FG material primarily lies in the significantly lower temperature-independent resistance, $R_{const.}$. Without magnetic field compensation, we are limited to a certain $R_{const.}$ in our test environment due to magnetic hygiene. 
An observed lower limit of approximately 1.8~n$\Omega$ for $R_{const.}$ is depicted in Fig. \ref{fig:Rres_vsdifflength}. Furthermore, this value is derived from our measurements at 1.5~K, which overestimates the true residual resistance value by a maximum of $\leq 1~\text{n}\Omega$.\\
Grain boundaries in polycrystalline materials typically enhance atomic diffusion, a characteristic less pronounced in LG than in FG. Hence, higher temperatures may be necessary for LG to achieve similar diffusion lengths and oxygen concentrations. Therefore, it is expected that the trend of $R_{const.}$ as a function of diffusion length for LG material, will be shifted downwards as can also be seen in Fig.~\ref{fig:Rres_vsdifflength}.\\
A question raised by the analysis presented here is the extent to which diffusion lengths of several hundred nanometers should impact RF performance when the London penetration depth $\lambda_L$ for niobium is only roughly 40~nm \cite{LondonPenetrationdepth_PhysRev.139.A1515}. 
The concentration relation in our parameter space $234~\text{nm}~<~l~<~2655~$nm from eq. \ref{eq:diffusionequation} is $C(\lambda_L) \approx 0.5 \cdot C(l)$. Concerning the above mentioned assumption of a finite source of oxygen at the niobium surface, we conclude that the diffusion length $l$ correlates with the total amount of oxygen within the RF penetration depth. Therefore, the RF performance is related to the diffusion length $l$. \\
%
%
In the case of the 3 hours 250°C treatment, the quality factors are lower and for 350°C they are similar to medium $l$ treatments but shifted to higher gradients with occurring HFQS which leads to the conclusion that the upper limit of the mid-T effect is reached.
%
%

\subsection{SIMS results from witness samples}

\begin{figure*}[!htb]
    \centering
        \subfigure[]{\includegraphics*[width=0.49\textwidth]{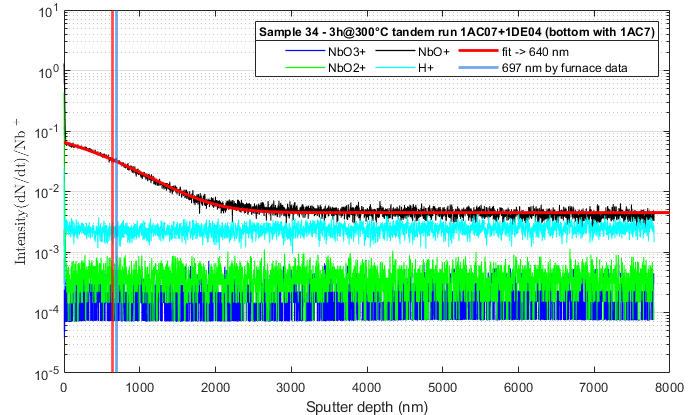}}
        \subfigure[]{\includegraphics*[width=0.49\textwidth]{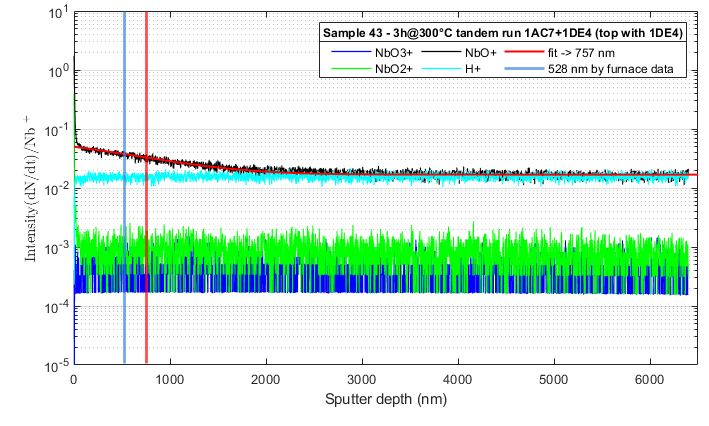}}
        
    	\subfigure[]{\includegraphics*[width=0.49\textwidth]{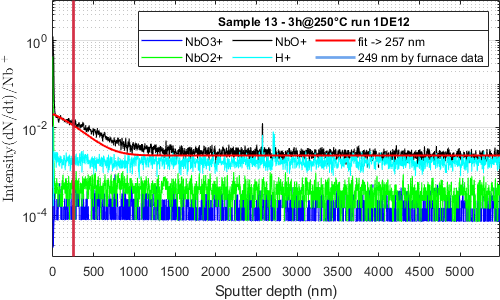}}
        \subfigure[]{\includegraphics*[width=0.49\textwidth]{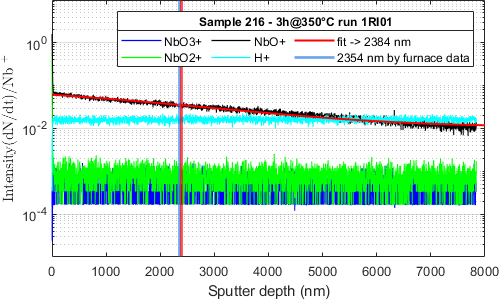}}

        \subfigure[]{\includegraphics*[width=0.49\textwidth]{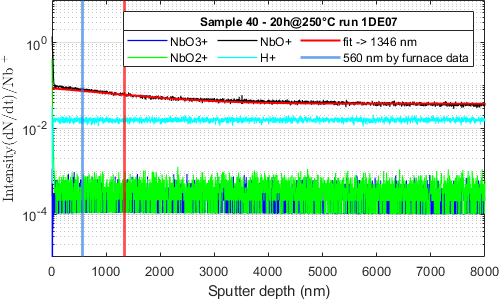}}
        \subfigure[]{\includegraphics*[width=0.49\textwidth]{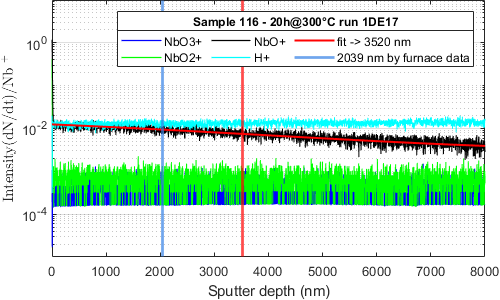}}
    \caption{ SIMS data of niobium samples treated with a cavity. Counts per second vs. depth are shown. The diffusion lengths, calculated from the temperature profiles (blue) as depicted in Fig. \ref{fig:Tprofilesall} and obtained through a nonlinear least squares fit to the SIMS data (red), are indicated by vertical lines.}
     \label{fig:SIMS_results}
    \end{figure*} 

Time-of-flight Secondary Ion Mass Spectrometry (TOF-SIMS) measurements were conducted on samples accompanying the respective cavities during the complete treatment procedure to compare the calculated diffusion lengths with experimental data. These samples underwent comparable pre-treatments like the cavities and were subsequently annealed together with them in the furnace. Measurements were conducted on a total of six differently treated samples. The TOF-SIMS analyses were conducted by using a TOF-SIMS instrument at Fraunhofer IFAM Bremen, Germany. Parameters included excitation with a 25~keV Bi liquid metal ion source in bunched mode, an analysis area of 50~x~50~$\text{µm}^2$, and charge compensation using a pulsed electron source. Sputter parameters involved 3~keV Argon ions, an area of 150~x~150~$\text{µm}^2$, and a partial pressure in the Ar source of $3\cdot10^{-5}$~mbar.\\
Fig.~\ref{fig:SIMS_results} shows profiles normalized to Nb+ obtained from SIMS analyses. The sputtering rates were determined by measuring the depth of the crater on the sputtered surface using a laser scanning confocal microscope, after each measurement. The TOF-SIMS depth profiles were recorded exclusively in the positive secondary ion polarity. As expected, the oxygen profile is particularly well reflected in the NbO+ signal. We aimed to obtain profiles at depths corresponding to the calculated oxygen diffusion lengths. Figures \ref{fig:SIMS_results} (a)-(f) depict profiles up to a depth of 8000~nm. 
A nonlinear least square fit was applied to the NbO+ signal according to the concentration equation
\begin{equation}
     \frac{C}{C_0} = erfc \left( \frac{z}{2\sqrt{Dt}} \right),
     \label{diffconc}
\end{equation}
where $C$ is the concentration, $C_0$ is the surface concentration and $z$ the depth into a semi-infinite medium with $z=0$ at the surface, derived from the solution to the second Fick's law for $z>0$ \cite{FickLaw}. From this, the diffusion length was determined using $l = \sqrt{Dt}$. The deviating results for the diffusion lengths from the fit to the NbO+ signal and those calculated from furnace temperature data are marked with vertical lines in the figures. For the 3~h 300°C tandem run of cavities 1AC07 and 1DE04, two samples were placed inside the recesses of the furnace frame near each cavity's top flange. The calculated diffusion length agrees very well with the value determined from the fit for one sample from the tandem run, as shown in Fig.~\ref{fig:SIMS_results}~(a). For the second sample shown in Fig.~\ref{fig:SIMS_results}~(b), a larger diffusion length with a deviation of 381~nm is the result of the fit. For samples from the 3~h 250°C Fig.~\ref{fig:SIMS_results}~(c) and 3~h 350°C Fig. \ref{fig:SIMS_results}~(d) treatments, the calculated and measured diffusion lengths again correspond very well. However, the values from extended treatments of 20~h at 250°C Fig. \ref{fig:SIMS_results}~(e) and 20~h at 300°C Fig. \ref{fig:SIMS_results}~(f) differ significantly, showing considerably larger diffusion lengths for the measured values.\\
The two factors, namely different oxygen concentrations and sputtering rates depending on the grain orientations and the number of grain boundaries, could account for these deviations. 
As previously demonstrated \cite{JAngle_10.1116/6.0001741}, the sputtering rate can also vary with respect to grain orientation, which would cause varieties in oxygen concentration.\\
In cases where the results for the diffusion lengths from the fits to the SIMS data deviate, the calculated values are consistently overestimated. This confirms that grain boundaries could be a significant cause of these discrepancies as they would be expected to lead to higher concentrations. Particularly with the small analysis area of only 50x50~µm$^2$, it is expected that the variation in the number of grain boundaries ranges from none to a few. Further measurements on single-crystal samples are planned and might clarify whether this is the main cause.



\section{Influence of Trapped Magnetic Flux on cavity performance}

While the mid-T heat treatment is beneficial for enhancing the intrinsic quality factor,
it also introduces concerns about increased sensitivity to trapped magnetic flux, leading to elevated surface resistance \cite{HIto}. The efficiency of flux expulsion during cool-down, crucial for cavity performance, is influenced by various factors, including geometry, niobium grain size, grain orientation, cool-down velocity, spatial temperature gradient, and magnetic flux density. In the following, the influence of a spatial temperature gradient on the cavity performance is especially analyzed. In addition, measurements of the sensitivity to trapped magnetic flux of mid-T heat treated cavities are introduced.

\subsection{Spatial temperature gradient}

\begin{figure}[!htb]
   \centering

    \subfigure[]{\includegraphics*[width=0.24\columnwidth]{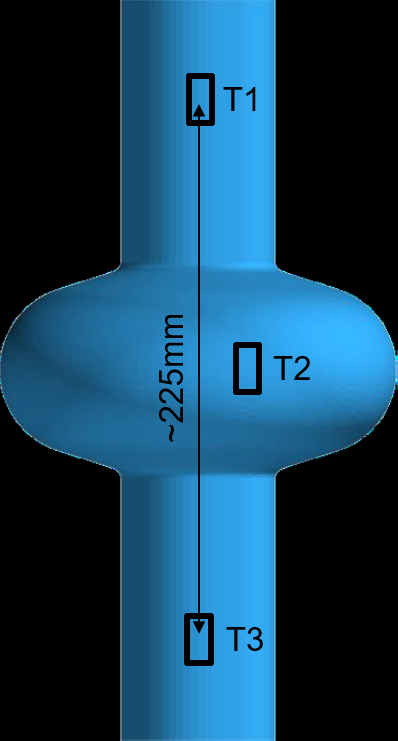}}
    \subfigure[]{\includegraphics*[width=0.74\columnwidth]{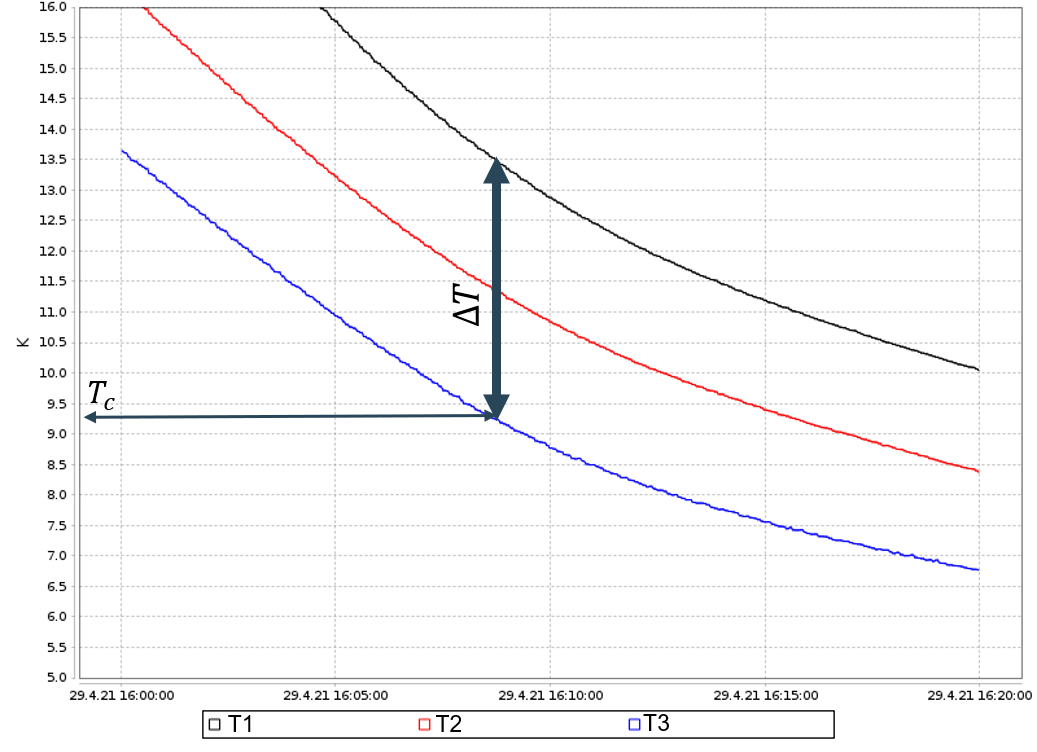}}
   
   \caption{Demonstration of the methodology employed to determine the spatial temperature gradient, utilizing temperature data measured by two sensors installed on the cavity.}
   \label{fig:spatialtempgradcalc}
\end{figure}

\begin{figure}[!htb]
   \centering
   \includegraphics*[width=1.0\columnwidth]{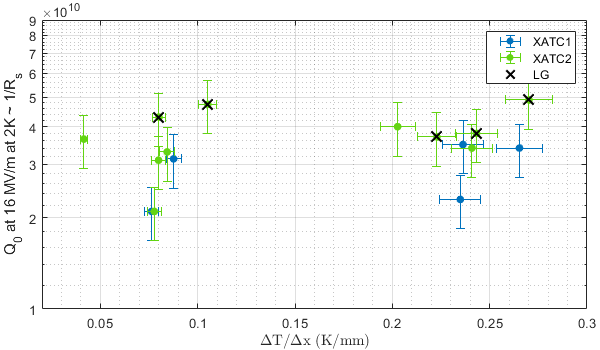}
   \caption{Measured $Q_0$ at 16 MV/m and at 2~K for all mid-T heat treatments against spatial cooldown gradient.}
   \label{fig:qmaxcooldown}
\end{figure}

\begin{figure}[!htb]
   \centering
   \includegraphics*[width=1.0\columnwidth]{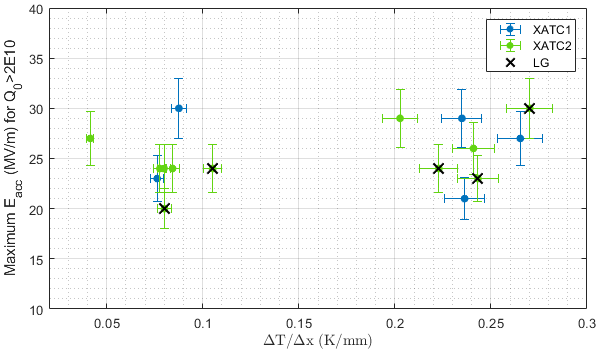}
   \caption{Maximal achieved gradient at 2~K (for $Q_0>2\cdot10^{10}$) for all mid-T heat treatments against spatial cooldown gradient.}
   \label{fig:Emaxcooldown}
\end{figure}

\begin{figure}[!htb]
   \centering
   \includegraphics*[width=1.0\columnwidth]{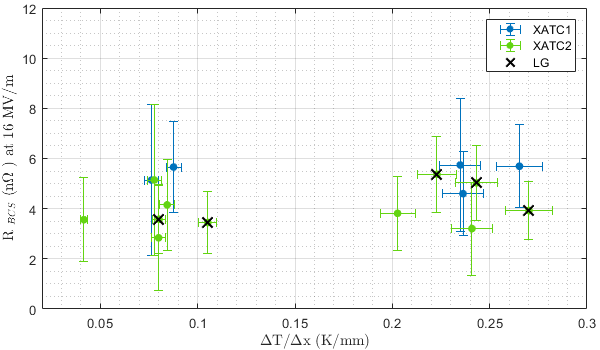}
   \caption{$R_{BCS,2K} \approx R_{S,2~K}-R_{S,1.5~K}$ for all mid-T heat treatments against spatial cooldown gradient.}
   \label{fig:Rbcscooldown}
\end{figure}

\begin{figure}[!htb]
   \centering
   \includegraphics*[width=1.0\columnwidth]{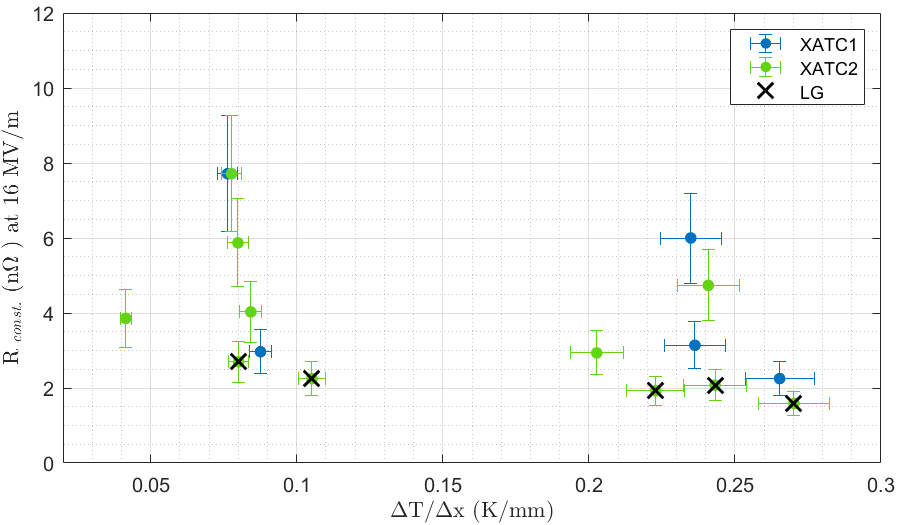}
   \caption{Surface resistance at 1.5~ K, nearly equivalent to the temperature independent resistance $R_{const.}$ for all mid-T heat treatments against spatial cooldown gradient.}
   \label{fig:Rrescooldown}
\end{figure}

To explore the spatial temperature gradient along the cavity during a cooldown, values of two temperature sensors are used, as illustrated in Fig.~\ref{fig:spatialtempgradcalc}. The temperature gradient is determined using
\begin{equation}
    \frac{\Delta T}{\Delta x} = \frac{T_3(9.2~\text{K})-T_1}{\Delta x = 225~\text{mm}},
    \label{eq:tempgrad}
\end{equation}
where $T_1$ represents the upper-temperature reading when $T_3$, the lower temperature sensor, reaches a value of $T_c~=~9.2$~K. The distance between the sensors is approximately $\Delta x~=~(225 \pm 10)~\text{mm}$.\\
To establish correlations with the cavity performance, the temperature gradient from eq. (\ref{eq:tempgrad}) is shown against $Q_0$ at 16~MV/m in Fig.~\ref{fig:qmaxcooldown}, and the maximum achieved accelerating gradient in Fig.~\ref{fig:Emaxcooldown} where XATC1\&2 denote our two cryostats. However, no correlations can be identified. Nevertheless, it appears that due to our two test cryostats and the specifically established cryogenic procedures, two distinct regions of cooling gradients exist. Although there are preferred cooling rates, it can be observed that they do not affect the quality factor. Subsequently, correlations with the $R_{BCS}$ in Fig.~\ref{fig:Rbcscooldown} and residual resistance in Fig.~\ref{fig:Rrescooldown} are investigated. Again, no correlations with the gradients are found. 

\subsection{Magnetometric mapping system to obtain the Sensitivity S}

\begin{figure}[htbp]
\centering
\includegraphics[width=1.0\columnwidth]{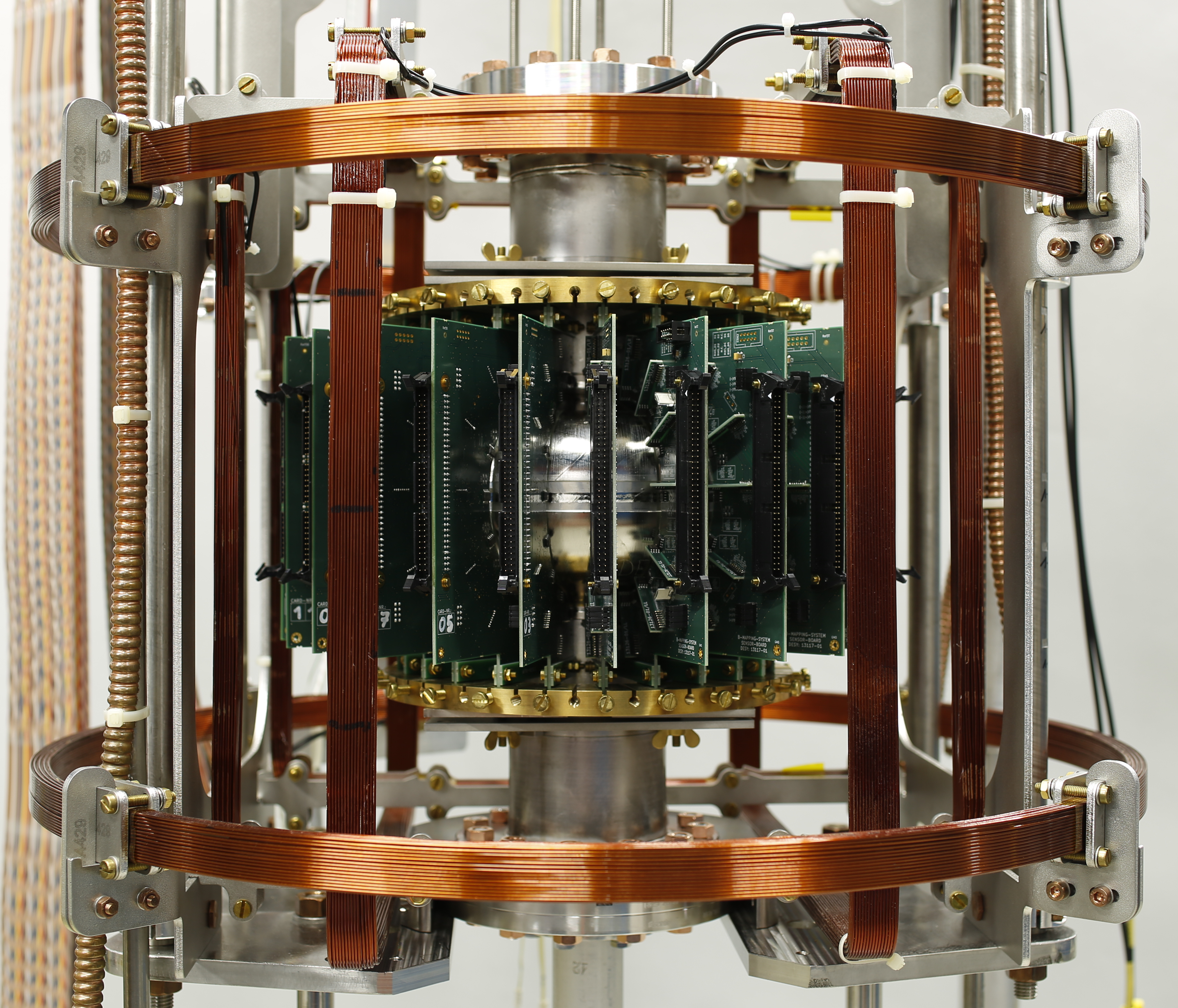}
\caption{Experimental setup: Cavity under test surrounded by the magnetometric mapping system before the assembly of the supply- and signal lines. The vertical Helmholtz coil is used to apply a defined magnetic flux density $B_{applied}$ during each cool down.}
\label{HC}
\end{figure}

\begin{figure}[htbp]
\centering
\includegraphics[width=1.0\columnwidth]{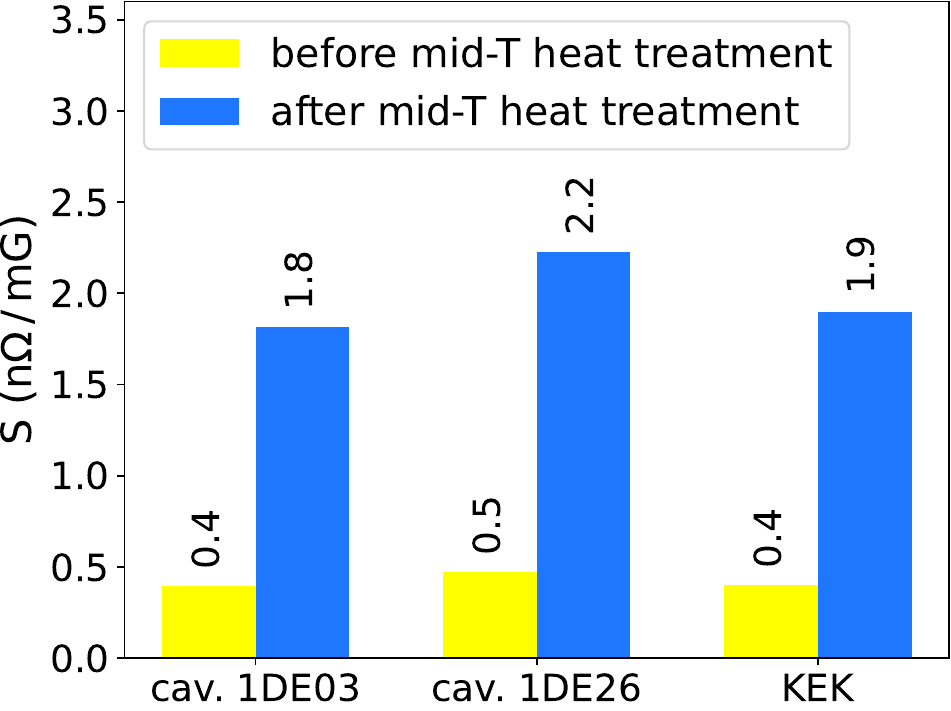}
\caption{Average sensitivity to trapped magnetic flux S of the fine-grain cavity 1DE03 and the large-grain cavity 1DE26 obtained by evaluating the interpolation function of $Q(E)$ at an accelerating gradient of \SI{4}{MV/m}. Results from KEK are added for comparison \cite{HIto}. Here the post mid-T heat treatment results are shown together with pre-treatment baseline data of a different cavity.}
\label{S}
\end{figure}

In this section, we investigate the impact of trapped magnetic flux after the mid-T heat treatment, employing a novel magnetometric mapping system.\\
%
Even though, $R_{BCS}$ decreases after the mid-T heat treatment, the total surface resistance $R_{S}$ given by
\begin{equation}
R_S(T, B) = R_{BCS}(T) + R_{res} + R_{flux}(B),
\end{equation}
may increase under the same test conditions due to a potentially growing contribution $R_{flux}$ linked to a grown so-called sensitivity S to trapped magnetic flux \cite{HIto} as observed during earlier studies \cite{SamMidT, HIto}.\\
$S$ is defined as the ratio
\begin{equation}
S = \frac{\Delta R_S}{B_{trap}},
\end{equation}
where $\Delta R_S$ represents the change of the surface resistance assumed to be linked to the fraction of the ambient flux trapped in the cavity cell $B_{trap}$.\\
During the DESY mid-T campaign, the change of S was exemplarily determined for the fine-grain cavity 1DE03 and the large-grain cavity 1DE26 before and after the 3-hour 300°C mid-T heat treatment. To calculate S, first $B_{trap}$ needs to be obtained for each cool down. For this purpose, the magnetometric mapping system introduced in \cite{wolff21,Wolff2023} and based on previous work at Helmholtz-Zentrum Berlin \cite{BSchmitz18,Kramer20} was used to consider possible differences in the flux expulsion behaviour alongside the cavity equator. $B_{trap}$ is determined by the test procedure described in \cite{Wolff2023} by utilizing the mean value of only the vertical component of the equator sensor group in a defined ambient field $B_{applied}$ of \SI{10}{µ T} applied by the vertical Helmholtz coil shown in Fig. \ref{HC}.
Contrary to the test procedure in \cite{Wolff2023}, a magnetic flux expulsion ratio of 1.53 for a cavity and a distance of the sensor center point to the cavity surface of \SI{6.5}{mm} in ideal Meissner state is used. This value of the flux expulsion ratio was obtained by an experimental method described in \cite{Huang18}.\\
Here the cavity under test is first cooled down below the critical temperature $T_{c}$ ideally in a zero field environment. Afterwards, a defined magnetic flux density is applied by the vertical Helmholtz coil and the magnetic flux density in superconducting state $B_{sc\_Meissner}$ is measured at the equator. Finally, the cavity is warmed up above $T_{c}$ to record the flux density in normal conducting state $B_{nc}$. The expulsion ratio in the ideal Meissner state is then given by $B_{sc\_Meissner}/B_{nc}$.\\
For both cavities, four independent tests were conducted before and after the mid-T heat treatment. Each of these tests consists of a cool down under varying cool down dynamics as described in \cite{Wolff2023} followed by a vertical performance test. $\Delta R_S$ is then obtained by means of an initial $Q_{0}$ vs. $E_{acc}$ curve recorded with no additional flux applied as an approximation of $R_{flux}$ as
\begin{equation}
\Delta R_S = R_S(B_{applied}) - R_{S\_initial} \approx R_{flux}.
\end{equation}
The average value for S is obtained by evaluating the interpolation function of $Q(E)$ at an accelerating gradient of \SI{4}{MV/m} and is given in Fig. \ref{S}.
S increased independently of fine or large-grain material by a factor of 4 to 5 after the mid-T heat treatment. In case of the fine-grain cavity 1DE03 from \SI{0.4}{n$\Omega$/mG} to \SI{1.8}{n$\Omega$/mG} and in case of the large-grain cavity 1DE26 from \SI{0.5}{n$\Omega$/mG} to \SI{2.2}{n$\Omega$/mG}. These results match with earlier studies conducted at KEK. There, S increased in a cavity overarching comparison from \SI{0.4}{n$\Omega$/mG} to \SI{1.9}{n$\Omega$/mG} after a \SI{300}{\degree C} for \SI{3}{h} mid-T heat treatment \cite{HIto}.

\section{Summary and Conclusions}

Efforts in refurbishing a niobium retort furnace lead to good prerequisites for a successful mid-T heat treatment campaign on 1.3~GHz single-cell TESLA-shaped cavities.
Through 19 treatments involving temperatures ranging from 250°C to 350°C and holding times from 3 to 20 hours, we have gained an understanding of the effective diffusion length and its impact on cavity performances.
The analysis of $Q_0$ versus $E_{acc}$ curves and the $R_{BCS}$ behaviour revealed distinct features characteristic of mid-T treated cavities, including an anti-Q slope with a maximum in $Q_0$ at around 16-20~MV/m, and a limited field strength of 25-35~MV/m with high reproducibility.\\
%
Grouping treatments based on their calculated diffusion lengths allows for a systematic examination of performance trends. Treatments with short $l$ demonstrate lower $Q_0$ performance with typical gradient limitation compared to those at medium $l$.
Treatments at long $l$ exhibit a similar $Q_0$ to medium $l$ and reveal unexpected high-field $Q$-slopes with the highest gradients observed.
All these results suggest that the treatments at short $l~\sim~250~nm$ and long $l~>~1700~nm$
are near the limits of the mid-T heat treatment range in which the typical characteristics can be observed.
%
Degradation in $Q_0$ after first quenching is observed in some cases without, until now, identified systematic cause. However, it can be cured by thermal cycling.
As the temperature independent resistance (at 16~MV/m) decreases with higher diffusion lengths and the BCS resistance (at 16~MV/m) increases at the same time, this leads to limited scope for parameter (temperature and baking time) optimization.
The range of the BCS resistance of our investigation is 3-$6~\text{n}\Omega$ and therefore significantly lower than the typical baseline performance.
The observed tendency emphasizes the critical role of diffusion length in determining cavity performance, with $l >> \lambda_L$ influencing the RF behaviour. However, the complexity of the process suggests that diffusion length may not be the sole determinant of cavity performance.
Our findings on LG cavities show lower temperature independent resistances but a BCS resistance in agreement with FG cavities. This result indicates that grain boundary diffusion is not responsible for the effect of the mid-T heat treatment. Otherwise for LG cavities the observed mid-T characteristics would not be expected in the investigated parameter range.
However, it is noteworthy that only the diffusion lengths between 500~nm and 1800~nm have been evaluated for LG cavities and more treatments beyond this range are under preparation. In the parameter range investigated here, gradients above 35~MV/m do not appear achievable for $Q_0>2\cdot10^{10}$, consistent with findings from other studies \cite{He_2021,HIto}. The here presented experimental work delivers no explanation for the gradient limitation.\\
SIMS measurements on samples experimentally confirm our calculated diffusion lengths in some cases, while in others they are overestimated. This discrepancy could be due to grain boundaries within the measurement area causing deviations. This will be further investigated using LG material.\\
Investigations of the spatial temperature gradient show no correlation with the performance of mid-T heat-treated cavities. The gradients varied between 0.04 and 0.27~K/mm and were not actively controlled during the measurement, but rather resulted from our predefined cryogenic cooling procedures.\\
%
%
Finally, we applied a recently developed B-mapping system to investigate the sensitivity to trapped magnetic flux. The study revealed an increase of the sensitivity in post-mid-T heat treatment for both FG and LG material by a factor of 4-5.
%
Although the temperature independent resistance of LG cavities was significantly lower than that made of FG material, no differences in sensitivity between the materials were detected. Differences in temperature independent resistance are evidently not only associated with sensitivity but are primarily related to the amount of flux that is trapped. However, it is possible that improved (lower) sensitivity to trapped flux at higher diffusion lengths could explain the observed trend of reduced temperature independent resistance with increasing diffusion lengths. Research into flux expulsion behaviour will be continued, including an investigation of its correlation with oxygen diffusion length and a possible dependency on the achievable field strength.\\
%
%
Future studies will focus on a set of 18 newly produced LG and FG single-cell cavities to explore the relationship between diffusion length and RF performance further. Additionally, comparisons with variations in time and temperature parameters will provide valuable insights into treatment optimization, especially achieving the same diffusion lengths with different treatment durations instead of different temperatures. Furthermore, the gaps between the various already studied diffusion lengths will be explored.
Promising parameters will be transferred to 9-cell cavities.\\
Overall, our findings contribute to the ongoing efforts to enhance the performance of SRF cavities for next-generation accelerator facilities.




\section{Acknowledgement}
We express our gratitude to the DESY and University of Hamburg SRF team for fruitful discussions and especially the cleanroom staff and the AMTF team for their essential support. Additionally, we appreciate 
the Fraunhofer IFAM in Bremen for the SIMS measurements. This work was supported by the Helmholtz Association within the topic Accelerator Research and Development (ARD) of the Matter and Technologies (MT) and by the European XFEL R\&D Program.
\printbibliography
\end{document}